\documentclass[letterpaper,titlepage,11pt]{article}
\usepackage{latexsym,amssymb,amstext,amsmath}
\usepackage{slashed}
\usepackage{mathrsfs}
\usepackage{color}
\usepackage{enumerate}
\usepackage{graphicx}
\usepackage{tikz}
\usepackage{latexsym}
\usepackage{cite}
\allowdisplaybreaks

\usepackage[
colorlinks=true,
linkcolor=blue,
urlcolor=red,
filecolor=green,
citecolor=red,
pdfstartview=FitV,
pdftitle={},
pdfauthor={Mehmet Ozkan},
pdfsubject={},
pdfkeywords={},
pdfpagemode=FullScreen,
bookmarksopen=true
]{hyperref}

\newcommand{\bea}{\setlength\arraycolsep{2pt} \begin{eqnarray}}
	\newcommand{\eea}{\end{eqnarray}}
\newcommand{\nn}{\nonumber}

\usepackage{hyperref}

\newsavebox{\uuunit}
\sbox{\uuunit}
{\setlength{\unitlength}{0.825em}
	\begin{picture}(0.6,0.7)
		\thinlines
		\put(0,0){\line(1,0){0.5}}
		\put(0.15,0){\line(0,1){0.7}}
		\put(0.35,0){\line(0,1){0.8}}
		\multiput(0.3,0.8)(-0.04,-0.02){12}{\rule{0.5pt}{0.5pt}}
		\end {picture}}

	\def\be{\begin{equation}}
		\def\ee{\end{equation}}
	\def\ba{\begin{array}}
		\def\ea{\end{array}}
	\def\bea{\begin{eqnarray}}
		\def\eea{\end{eqnarray}}
	\def\bd{\begin{displaymath}}
		\def\ed{\end{displaymath}}
	
	\def\nn{\nonumber}
	
	\def\a{\alpha}

	\def\d{\delta}
	
	\def\e{\epsilon}

	\def\k{\kappa}
	\def\l{\lambda}
	
	\def\m{\mu}
	\def\n{\nu}
	\def\r{\rho}
	\def\s{\sigma}
	
	\def\t{\tau}

	\def\o{\omega}
	\def\O{\Omega}

	\def\nn{\nonumber}

	\def\cL{\mathcal{L}}

	\makeatletter
	\@addtoreset{equation}{section}
	\makeatother
	
\begin{document}
		\begin{titlepage}
			
			\bigskip
			\begin{center}
				{\LARGE \bfseries  Non-Relativistic and Ultra-Relativistic Scaling Limits of Multimetric Gravity}
				\\[10mm]
				%
				%
				%
				%
				%
				%
				%
				
				\renewcommand{\thefootnote}{\alph{footnote}}
				{\large Ertu\u{g}rul Ekiz$^{~1}$}\footnote{Email: {\tt ekize15@itu.edu.tr}},
				{\large Oguzhan Kasikci$^{~1}$}\footnote{Email: {\tt kasikcio@itu.edu.tr}},
				{\large Mehmet Ozkan$^{~1}$}\footnote{Email: {\tt ozkanmehm@itu.edu.tr}},
				{\large Cemal Berfu Senisik$^{~1}$}\footnote{Email: {\tt senisik@itu.edu.tr}} \\
				and  {\large Utku Zorba$^{~2}$}\footnote{Email: {\tt utku.zorba@boun.edu.tr}}
				
				\setcounter{footnote}{0}
				\renewcommand{\thefootnote}{\arabic{footnote}}
				
				\vspace{0.5cm}

				${}^1$	{\em  \hskip -.1truecm Department of Physics, Istanbul Technical University,  \\
					Maslak 34469 Istanbul, Turkey  }\\
				\vskip .2truecm
				
				${}^{2}$ {\em  \hskip -.1truecm Physics Department, Boğaziçi University,  \\
					34342 Bebek, Istanbul, Turkey  }

				\vspace{1.8cm}

			\end{center}
			
			\vspace{3ex}

			\begin{center}
				{\bfseries Abstract}
			\end{center}
			\begin{quotation} \noindent
				
				We present a method of contraction that can be applied to re-construct the recent extended non-relativistic and ultra-relativistic algebras as well as corresponding action principles. The methodology involves the use of multiple copies of Poincar\'e  algebra. Consequently, the contraction defines non-relativistic or ultra-relativistic limits of multimetric theories of gravity. In particular, we show that the non-relativistic scaling limit of bi-metric gravity corresponds to the recent formulation of an action principle for Newtonian gravity with a constant background mass density.

			\end{quotation}
			
			\vfill
			
		\end{titlepage}
		\setcounter{page}{1}
		\tableofcontents
		
		\newpage

		\section{Introduction}
		\paragraph{}
		
		
		Lie algebra expansion \cite{Hatsuda:2001pp,deAzcarraga:2002xi,deAzcarraga:2007et,Bergshoeff:2019ctr} is a powerful tool to generate interesting gravitational theories starting from the first order formulation of the (cosmological) Einstein-Hilbert action. On the one hand, massive gravity theories that are consistent with the holographic c-theorem have been shown to arise from the truncation of an infinite-dimensional Lie algebra that is closely connected to the Lie algebra expansion of the $\rm AdS$ algebra and the cosmological Einstein-Hilbert action \cite{Bergshoeff:2021tbz}. On the other hand,  this procedure was the main tool to construct the action principle for Newtonian gravity in  first-order formulation (see \cite{VandenBleeken:2017rij,Hansen:2019pkl} for its second order formulation and the relevant $1/c^2$ expansion) as well as establishing new, extended, two and three-dimensional non/ultra-relativistic gravity models \cite{Papageorgiou:2009zc, Bergshoeff:2016lwr,Hartong:2016yrf,Aviles:2018jzw, Ozdemir:2019orp, deAzcarraga:2019mdn,Concha:2019lhn, Penafiel:2019czp, Gomis:2019fdh, Ozdemir:2019tby, Gomis:2019sqv, Gomis:2019nih, Kasikci:2020qsj, Concha:2020sjt, Concha:2020ebl, Concha:2020eam, Concha:2020tqx, Concha:2021jos, Gomis:2022spp,  Grumiller2020, Gomis2020, Ravera:2022buz, Concha:2022you, Concha:2021llq, Ravera:2019ize, Ali:2019jjp, Concha:2021jnn}.
		In fact, the massive gravity models of \cite{Bergshoeff:2021tbz} arise as scaling limits of ghost-free bi-gravity models \cite{Paulos:2012xe,Afshar:2014dta} which later lead to the discovery that there exist trajectories in the parameter space of bi-gravity theories that connect the central charges of bulk/boundary unitary three-dimensional bi-gravity models to non-unitary massive gravity theories by a continuous change of scaling parameter \cite{Bergshoeff:2013xma,Ozkan:2019iga,Sevim:2019scg}. It is, thus, a natural question whether one can unify the Lie algebra expansion and the scaling limit together to define a non/ultra-relativistic limit for bimetric  and multimetric models of gravity to establish similar connections between physical quantities. In this paper, we will show that this is indeed the case by presenting a systematic procedure that relates the space-time decomposed multimetric gravity to extended non-relativistic and ultra-relativistic models of gravity.

		
		In building extended non/ultra-relativistic gravity, the main motivation comes from the formulation of an action principle for Newtonian gravity. This construction requires one to go beyond the standard Bargmann algebra by an extension with additional three new generators. This extension, which was originally formulated by $1/c^2$ expansion of Einstein-Hilbert action \cite{Hansen:2019pkl}, separates the strong gravitational effects from the relativistic effects  and it has been shown that in the presence of a \textit{twistless torsion}, the Newtonian gravity action can successfully explain  the three classical tests of general relativity \cite{VandenBleeken:2019gqa, Ergen:2020yop, Hansen:2019vqf, Hansen:2020pqs}. Thus, although the starting point was the $1/c^2$-expansion of the Einstein gravity, the models that arise in each order in the expansion are novel both theoretically and phenomenologically, and can be studied in their own right.
		
		The main procedure to find extended non/ultra-relativistic algebras and corresponding gravity models is the Lie algebra expansion. It is based on the splitting of the generators of a Lie algebra into even and odd classes followed their series expansion with respect to the class that they belong to. If the algebra under consideration is chosen to be the space-time split Poincar\'e algebra, then the expansion yields either extended non-relativistic or extended ultra-relativistic algebras \cite{Bergshoeff:2019ctr}. The corresponding gravity models can also be found in the same spirit, that is, one can start with the first-order formulation of the Einstein-Hilbert action, perform the space-time splitting and expand the fields in accordance with their corresponding generator \cite{Bergshoeff:2019ctr}. This procedure successfully gives rise to the action principle for the Newtonian gravity (see Appendix \ref{AppA} for the equivalence of the first and the second order formulations) and various two and three-dimensional models have been constructed/reconstructed in the recent literature. The  transformation rules for the matter fields can also be found by this methodology \cite{Kasikci:2021atn} (see \cite{Hansen:2020pqs} for the corresponding $1/c^2$-expansion) including the rigid supermultiplets of extended superalgebras. The local transformation rules for supersymmetric models is an open problem to date. In particular, in three-dimensions, the nature of expansion does not  even  allow for the rigid supermultiplets for the algebras where the supergravity actions can be written \cite{Kasikci:2021atn}.
		
		In a separate development, it has been found that the massive gravity models that admit the holographic c-theorem arise from the truncation of an infinite dimensional Lie algebra and its corresponding gauge theory of gravity \cite{Bergshoeff:2021tbz}. This infinite-dimensional Lie algebra is a cosmological algebra and if the cosmological parameter is set to zero, it simply becomes the Lie algebra expansion of the $D$-dimensional Poincar\'e algebra. The inclusion of the cosmological constant is equivalent to the infinite-dimensional expansion of the $\rm (A)dS$ algebra given the fact that the cosmological constant is scaled with the expansion parameter, i.e. $\Lambda \to \Lambda/\lambda^2$ where $\Lambda$ is the cosmological constant and $\lambda$ represents the expansion parameter. The truncation of the resulting infinite-dimensional Lie algebra gives rise to gravity models with a set of auxiliary fields, which, when solved and substituted back into the action, becomes massive gravity models that are compatible with the holographic c-theorem. These models include the new massive gravity \cite{Bergshoeff:2009hq} and its various extensions \cite{Sinha:2010ai,Paulos:2010ke}, all of which have been shown to be related to multimetric  gravity by means of a scaling limit \cite{Paulos:2012xe,Afshar:2014dta}.
		
		In this paper, we investigate the connection between these two seemingly unrelated subjects. In particular, we have shown that the action principles for non/ultra-relativistic gravity models arise as a scaling limit of multimetric gravity. This can be thought as an important stepping stone towards an understanding of the solutions and the phenomenological aspects of non/ultra-relativistic gravity theories by performing the scaling limit that we discuss in this paper. We begin in Section \ref{AlgAndAct} by reminding the reader about the basics of the Lie algebra expansion and present the general formulation of extended non-relativistic and ultra-relativistic actions, giving particular attention to three and four dimensions. In Section \ref{MultiGrav}, we show that the non-relativistic gravity models with larger symmetries are scaling limits of multimetric  theories with a Lorentzian signature. This point is one of our key results, so let us be more precise with our statement. As mentioned, there is a direct connection between the Lie algebra expanded (A)dS algebra and the massive theories of gravity. When the cosmological constant is set to zero, a model that comes from a consistent truncation of the infinite-dimensional algebra does not describe massive gravity but it is a theory of gravity that is coupled to a set of gauge fields \cite{Bergshoeff:2021tbz}. These models arise from the scaling limit of a multimetric  theory in the absence of potential terms for the vielbein. Nevertheless, as they are directly relevant to the expansion of the Poincar\'e algebra, we first relate the multimetric  models with no potential to non-relativistic and ultra-relativistic gravity models. For example, the scaling limit of a bimetric gravity without potential terms is the gauge theory formulation of Newtonian gravity with no source. Based on our result for how to take the scaling limit, we then turn on the potential terms and establish their contribution to the non-relativistic and ultra-relativistic models. In the case of bi-gravity, the potential terms give rise to a constant background mass density for non-relativistic gravity. In Section \ref{URChapter}, we show that there is an analog construction for the ultra-relativistic gravity models. We show that ultra-relativistic gravity with extended symmetries also arises as a different limit of the same multi-gravity models, which resembles the Galilei / Carroll limits of General Relativity.
		We give our comments and conclusions in Section \ref{Discussion}.

		
		\section{Algebras and Actions}\label{AlgAndAct}
		\paragraph{}
		The Lie algebra expansion is a method to generate higher-dimensional Lie algebras starting from a lower-dimensional core Lie algebra. As we will discuss the details momentarily, it is based on a series expansion of Maurer-Cartan one-forms of the dual algebra. Thus, the expansion that generates larger Lie algebras also generates action principles by expanding a core action that is invariant under the core Lie algebra. In particular, for the space-time decomposed Poincar\'e (or (A)dS) algebra and the corresponding first-order formulation of the (cosmological)  Einstein-Hilbert action, the expansion yields non/ultra-relativistic gravity models at each order in expansion parameter \cite{Bergshoeff:2019ctr}. For example,
		\begin{eqnarray}
			\cL_{\rm GR} &=& \l \cL_{1} + \l^3 \cL_{3} + \l^5 \cL_{5} + \ldots \,,
			\label{Structure}
		\end{eqnarray}
		is the structure of the non-relativistic expansion where $\lambda$ is the expansion parameter and $\cL_n$ represents the Lagrangian at the relevant $\lambda^{n}$ order \textcolor{red}{\cite{Hatsuda:2001pp,deAzcarraga:2002xi,deAzcarraga:2007et,Bergshoeff:2019ctr}}. Note that each of these actions is invariant under the corresponding order of the expanded core Lie algebra. As can be seen from the structure of the expanded Lagrangian, the expansion and the scaling limit $(\lambda \to 0)$ yield the same result at the lowest order. For instance, in the case of \eqref{Structure}, the lowest order Lagrangian in the expansion, $\cL_1$, can also be found by the same expansion of the gauge fields, then rescaling the core Lagrangian $\cL_{\rm core}$ by a factor of $\lambda^{-1}$ and finally taking the limit $\lambda \to 0$ in which case the coefficients of all $\cL_n$ with $n >1$ vanishes. However, it is not possible to single out a Lagrangian $\cL_n$ with $n>1$ in this way as rescaling the core Lagrangian with $\lambda^{-n}$  would yield divergences in the coefficients of lower order terms. One way to isolate a higher-order Lagrangian to perform the scaling limit is to consider multiple copies of the same core algebra and combine the core Lagrangians to cancel out any lower-order terms that would cause divergences. In the case of Poincar\'e algebra, this means that we must consider multiple copies of Einstein-Hilbert action to obtain a proper scaling limit. This is the leading technical notion of this present paper, which then describes the non/ultra-relativistic scaling limits of multi-gravity models.
		Thus, this section is aimed to discuss the non-relativistic and ultra-relativistic expansions of the Poincar\'e algebra and Einstein-Hilbert action in first-order formulation to set the stage for multimetric  models and their non/ultra-relativistic scaling limits.
		
		\subsection{Lie Algebra Expansion and the Poincar\'e Algebra}
		\paragraph{}
		The Lie algebra expansion is a method that takes a core Lie algebra $\mathfrak{g}$ and produces new, higher dimensional algebras as long as $\mathfrak{g}$ can be written as a direct sum of two subspaces $V_0$ and $V_1$ that satisfies the following relations
		\begin{align}
			\left[V_0, V_0 \right] & \subset V_0 \,, & \left[V_0, V_1 \right] & \subset V_1 \,, & \left[V_1, V_1 \right] & \subset V_0 \,.
		\end{align} 
		Based on these relations, $V_0$ represent the even class of generators while $V_1$ represents the odd class. The direct sum structure of the core Lie algebra suggests that we may also assign a gauge field to each of the generators
		\begin{eqnarray}
			A_\mu  = A_\m^i X_i + A_\m^\alpha Y_\alpha \,, 
		\end{eqnarray} 
		where $X^i$ represents the even subset of generators while $Y^\alpha$ represents the odd ones. In the next step, we expand the gauge fields with an expansion parameter $\lambda$ with respect to the class that they belong to
		\begin{align}
			A^i & = \sum_{n=0}^{N_0} \lambda^{2n} A^i_{(2n)} \,, & A^\alpha & = \sum_{n=0}^{N_1} \lambda^{2n+1} A^\alpha_{(2n+1)} \,.
			\label{ExpansionCore}
		\end{align}
		Here, the sum can be extended to infinity to produce and infinite dimensional Lie algebra. The consistent truncation at order $g=(N_0,N_1)$ requires that either $N_0 = N_1$ or $N_0 = N_1 + 1$ is satisfied. With this expansion in hand, one can start with the Maurer-Cartan equations of the core Lie algebra  $\mathfrak{g}$, expand the gauge fields with respect to \eqref{ExpansionCore} and read off the structure constants of the expanded algebra from the expanded  Maurer-Cartan equations at each order. Equivalently, based on their even/odd character, we may expand the generators $\{X^i, Y^\alpha\}$ with $X_i \in V_0$ and $Y_\alpha \in V_1$ as follows
		\begin{align}
			X_i^{(2n)} & = \lambda^{2n} \otimes X_i\,, & Y_\alpha^{(2n+1)} &= \lambda^{2n+1} \otimes Y_\alpha \,.
			\label{GeneratorExpansion}
		\end{align}
		Then, using the commutation relations of the core algebra
		\begin{align}
			\left[X_i, X_j \right] & = f_{ij}{}^k X_k \,, & \left[X_i, Y_\alpha \right] & = f_{i\alpha}{}^\beta Y_\beta \,, & \left[Y_\alpha, Y_\beta\right] & = f_{\alpha \beta}{}^i X_i \,,
		\end{align}
		we may give the commutation relations for the expanded algebra as \cite{Gomis:2019nih} 
		\begin{align}
			\left[X_i^{(2m)}, X_j^{(2n)} \right] & = f_{ij}{}^k X_k^{(2m+2n)}\,, &  \left[X_i^{(2m)}, Y_\alpha^{(2n+1)} \right] & = f_{i\alpha}{}^\beta Y_\beta^{(2m+2n+1)}  \,,\nn\\\
			\left[Y_\alpha^{(2m+1)}, Y_\beta^{(2n+1)}\right] & = f_{\alpha \beta}{}^i X_i^{(2m+2n+2)} \,.
			\label{GenExp}
		\end{align}
		Note that the Jacobi identities for algebras are closed at each order without referring to higher-order commutators. Consequently, the group theoretical curvatures of the resulting truncated algebras satisfy the Bianchi identity without referring to higher-order terms \cite{Hatsuda:2001pp,deAzcarraga:2002xi,deAzcarraga:2007et,Bergshoeff:2019ctr}. With this result in hand, let us now turn our attention to the space-time split Poincar\'e algebra and its non/ultra-relativistic Lie algebra expansions. The $D$-dimensional Poincar\'e algebra consists of translations $(P_A)$ and Lorentz transformations $M_{AB}$ with the following non-vanishing commutation relations
		\begin{align}
			\left[M_{AB}, P_C \right] & = 2 \eta_{C[B} P_{A]} \,, & \left[M_{AB}, M_{CD}\right] & = 4 \eta_{[A[C} M_{D]B]} \,.
		\end{align}
		The space-time decomposition can be achieved by decomposing the $D$-dimensional index $A$ as $A= (0,a)$ in which case the generators are split as
		\begin{align}
			M_{AB} & = \{M_{0a} \equiv G_a, J_{ab} \}\,, & P_a & = \{P_0 \equiv H, P_a \} \,.
			\label{DecompGen}
		\end{align}
		In this case, the Poincar\'e algebra decomposes as
		\begin{align}
			\left[G_a, P_b\right] &= \d_{a b} H\,, &  \left[G_a, H\right] & = P_a \,, &\left[J_{a b}, P_c\right] &= \d_{b c} P_a - \d_{a c} P_b\,, \nn\\
			\left[J_{a b}, G_{c}\right] &= \d_{b c} G_a - \d_{a c} G_b \,, &\left[J_{a b}, J_{c d}\right] &= 4 \d_{[a[c}J_{d]b]}\,, & \left[G_a, G_b\right] & = J_{a b}\,. 
			\label{decomposealgebra}
		\end{align}
		Based on the spacetime decomposed Poincar\'e algebra, we may discuss the non-relativistic and ultra-relativistic Lie algebra expansions and action principles.
		
		\subsection{Non-Relativistic Algebras and Actions}\label{NRPrelim}
		\paragraph{}
		The non-relativistic higher-dimensional algebras are achieved with the following choice for the generators \cite{Bergshoeff:2019ctr}
		\begin{align}
			V_0 & = \{J_{ab}, H\}\,, & V_1 & = \{P_a, G_a\} \,.
		\end{align}
		With this choice of generators, we can follow the prescription that we presented in \eqref{GenExp}. Thus, the Lie algebra expansion of the spacetime decomposed Poincar\'e algebra is given by
		\begin{align}
			\left[G_a^{(2m+1)}, P_b^{(2n+1)}\right] &= \d_{a b} H^{(2m+2n+2)}\,, &  \left[G_a^{(2m+1)}, H^{(2n)}\right] & = P_a^{(2m+2n+1)} \,,  \nn\\
			\left[J_{a b}^{(2m)}, P_c^{(2n+1)}\right] &=2 \d_{c[b} P_{a]}^{(2m + 2n+1)} \,, & \left[J_{a b}^{(2m)}, G_c^{(2n+1)}\right] &=2 \d_{c[b} G_{a]}^{(2m + 2n+1)}  \,, \nn\\
			\left[J_{a b}^{(2m)}, J_{c d}^{(2n)}\right] &= 4 \d_{[a[c}J_{d]b]}^{(2m+2n)}\,, & \left[G_a^{(2m+1)}, G_b^{(2n+1)}\right] & = J_{a b}^{(2m+2n+2)}\,. 
			\label{NRInfinite}
		\end{align}
		To provide the well-known non-relativistic algebras that arise as a consistent truncation of this infinite-dimensional algebra, let's first focus on the simplest case where we have two even and two odd generators, $P_a^{(1)}, G_a^{(1)}, H^{(0)}, J_{ab}^{(0)}$. These generators satisfy the Galilei algebra
		\begin{align}
			\left[G_a, H\right] & = P_a \,,  &\left[J_{a b}, P_c\right] &= \d_{b c} P_a - \d_{a c} P_b\,, \nn\\
			\left[J_{a b}, G_{c}\right] &= \d_{b c} G_a - \d_{a c} G_b \,, &\left[J_{a b}, J_{c d}\right] &= 4 \d_{[a[c}J_{d]b]}\,,
			\label{GalAlb}
		\end{align}
		where we relabeled the generators as $P_a^{(1)} = P_a, G_a^{(1)} = G_a, H^{(0)} = H$ and $J_{ab}^{(0)} = J_{ab}$. According to the consistent truncation conditions, we may now add two more generators that belong to $V_0$, i.e., $H^{(2)}$ and $J_{ab}^{(2)}$ \footnote{The algebra with these extra two generators is known as the extended Bargmann algebra, which extends the Bargmann algebra with a new generator $J_{ab}^{(2)} = S_{ab}$ if $H^{(2)}$ is identified as the mass generator $ H^{(2)} \equiv M$. The Lie algebra expansion skips these two cases and directly goes from the Galilei algebra to the Newtonian algebra in four and higher dimensions \cite{Bergshoeff:2019ctr}.}. However, as we will discuss momentarily, this truncation does not have a corresponding invariant action principle that can be achieved by the expansion of Einstein-Hilbert action for in four and higher dimensions. The next consistent truncation requires two more additional generators of odd character, $P_a^{(3)}$ and $G_a^{(3)}$, in which case the algebra becomes identical to the one that underlies the first-order formulation of Newtonian gravity \cite{Hansen:2019vqf}
		\begin{align}
			\left[G_a, H\right] & = P_a \,,  &\left[J_{a b}, P_c\right] &= \d_{b c} P_a - \d_{a c} P_b\,, & \left[J_{a b}, G_{c}\right] &= \d_{b c} G_a - \d_{a c} G_b \,, \nn\\
			\left[J_{a b}, T_c\right] &= \d_{b c} T_a - \d_{a c} T_b\,, & \left[J_{a b}, B_c\right] &= \d_{b c} B_a - \d_{a c} B_b\,, & \left[J_{a b}, J_{c d}\right] &= 4 \d_{[a[c}J_{d]b]}\,, \nn\\
			\left[J_{a b}, S_{c d}\right] &= 4 \d_{[a[c}S_{d]b]}\,, & 	\left[G_a, P_b\right] &= \d_{a b} M \,, & \left[B_a, H\right] & = T_a  \,,\nn\\
			\left[G_a, M\right] & = T_a \,, &  \left[G_a, G_b\right] & = S_{a b}\,, & \left[S_{a b}, P_c\right] &= \d_{b c} T_a - \d_{a c} T_b \,,\nn\\
			\left[S_{a b}, G_c\right] &= \d_{b c} B_a - \d_{a c} B_b\,,
			\label{NewtAlg}
		\end{align} 
		where we labeled $H^{(2)} = M$, $J_{ab}^{(2)} = S_{ab}, P_a^{(3)} = T_a$ and $G_a^{(3)} = B_a$. 
		For the construction of an action principle, we first need to space-time decompose the gauge fields of the Poincar\'e algebra, namely vielbein $E_\mu{}^A$ and the spin connection $\Omega_\mu{}^{AB}$, i.e.
		\begin{align}
			E^A & = \{E^0 = T, E^a\}\,, & \Omega^{AB} & = \{\Omega^{0a} = \Omega^a, \Omega^{ab}\} \,.
			\label{DecompFields}
		\end{align}
		This step can then be  followed by their expansion in line with the expansion of their corresponding generator \eqref{DecompGen}. 
		\begin{align}
			T & = \sum_{n=0}^{N_0} \lambda^{2n} \tau_{(2n)} \,, & E^a & = \sum_{n=0}^{N_1}  \lambda^{2n+1} e^a_{(2n+1)} \,,\nn\\
			\Omega^{ab}  & = \sum_{n=0}^{N_0} \lambda^{2n} 	\omega^{ab}_{(2n)} \,, &  \Omega^a & = \sum_{n=0}^{N_1}  \lambda^{2n+1} \omega^a_{(2n+1)} \,.
			\label{ExpGF}
		\end{align}
		These expressions can be used in the space-time decomposed Einstein-Hilbert action in the first order formulation to generate invariant non-relativistic gravity models. To perform the expansion explicitly, let us focus on three and four dimensions, however our arguments are dimension independent. In four dimensions, the space-time decomposed action is given by
		\begin{eqnarray}
			\cL_{EH} &=& \epsilon_{ABCD}R^{AB}\wedge E^{C}\wedge E^{D} = 2\epsilon_{a b c} \left(R(\Omega^{a})\wedge E^b \wedge E^c - R(\Omega^{a b})\wedge E^c \wedge T \right) \,.
		\end{eqnarray}
		Here $R^{AB}$ refers to the group-theoretical curvature of the spin connection and $R(\O^a)$ and $R(\O^{ab})$ refer to its spacetime decomposition with respect to \eqref{DecompFields}. While these curvatures can simply be read off from the Poincar\'e algebra, it is useful to present their expression explicitly for future purposes
		\begin{align}
			R(\Omega^a) &= d \Omega^a + \Omega^{ab}\wedge \Omega_b\,, &	R(\Omega^{ab}) &= d\Omega^{ab} - \Omega^{ac} \wedge \Omega^b{}_c - \Omega^a \wedge \Omega^b \,.
			\label{DecompCurvatures}
		\end{align}
		The expanded action is then given by
		\begin{eqnarray}
			\cL &=& -2 \e_{abc} \sum_{m=0}^{N_0} \sum_{n=0}^{N_1} \sum_{\ell=0}^{N_0} \lambda^{2m+2n+2\ell +1}  R(\omega^{ab}_{(2m)}) \wedge e^c_{(2n+1)} \wedge \tau_{(2\ell)} \nn\\
			&& + 2 \e_{abc} \sum_{m=0}^{N_1} \sum_{n=0}^{N_1} \sum_{\ell=0}^{N_1} \lambda^{2m+2n+2\ell +3} R(\omega^a_{(2m+1)}) \wedge  e^b_{(2n+1)} \wedge  e^c_{(2\ell +1)}  \,.
			\label{4dNRExp}
		\end{eqnarray}
		As mentioned, the consistent truncation of the algebra as well as the action requires a certain relation between $N_0$ and $N_1$. In the case of the algebra, the necessary condition was that either $N_0 = N_1$ or $N_0 = N_1 +1$ must be satisfied. In the case of the action, the situation is more subtle. For example, let us consider the expansion to order $(N_0,N_1) = (1,1)$
		\begin{eqnarray}
			\cL &=& \lambda \left(-2 \e_{abc} R^{ab}(\o) \wedge e^a \wedge \tau \right) \nn\\
			&& + \lambda^3 \Big( -2 \e_{abc} \left(R^{ab}(\o) \wedge e^c \wedge m + R^{ab}(s) \wedge e^c \wedge \tau + R^{ab}(\o) \wedge t^c \wedge \tau  \right)\nn\\
			&& \qquad \,\,\,  + 2 \e_{abc} R^a(\O) \wedge e^b \wedge e^c  \Big)  \,,
		\end{eqnarray} 
		where we set
		\begin{align}
			\tau_{(0)} & = \tau \,, & \tau_{(2)} & = m \,, & \omega^{ab}_{(0)} &= \omega^{ab} \,, & \omega^{ab}_{(2)} & = s^{ab} \,,\nn\\
			e^a_{(1)} &= e^a \,, & e^a_{(3)} &= t^a \,, &  \omega^a_{(1)} & = \Omega^a &  \omega^a_{(3)} & = b^a \,. 
			\label{NewtonExpand}
		\end{align}
		Furthermore, the group theoretical curvatures $R^{ab}(\omega), R^{ab}(s)$ and $R^a(\O)$ can easily be read off by expanding the curvatures \eqref{DecompCurvatures} to the $\lambda^3$ order
		\begin{align}
			R^{ab}(\omega) &= d\o^{ab} - \o^{ac} \wedge \o^b{}_c \,, & R^{ab}(s) & = ds^{ab} - 2 \o^{ac} \wedge s^b{}_c - \o^a \wedge \o^b \,,\nn\\
			R^{a}(\o) & =  d\o^a + \o^{ab} \wedge \o_b \,.
				\label{NewtonianCurvatures}
		\end{align}
		In this Lagrangian, the $\lambda^1$-order model represents the Galilei gravity which is invariant under the Galilei algebra \eqref{GalAlb}. The $\lambda^3$-order Lagrangian is the first-order formulation of the Newtonian gravity which is invariant under \eqref{NewtAlg}. While we truncated the algebra and the expansion at order $(N_0,N_1) = (1,1)$, we could have stopped at order $(N_0,N_1) = (1,0)$. When this happens, the $\lambda$-order Lagrangian remains unaltered but the $\lambda^3$-order does not have the $R^{ab}(\o) \wedge t^a \wedge \tau$ term in Lagrangian which spoils its exact gauge invariance without referring to next-order terms. Although we provide a four-dimensional example here, the argument holds in general. It is better to think about the expansion as a truncation of the infinite-dimensional algebra and the action. Unless all terms that contribute to a certain $\lambda$-order are taken into account, the action is not gauge-invariant \cite{Bergshoeff:2019ctr}. In four-dimensions, the expanded Lagrangian \eqref{4dNRExp} indicates that the consistent truncation requires $N_0 = N_1$.  
		
		In three-dimensions, for instance, the argument we provide above means that the consistent truncation occurs if $N_0 = N_1 +1$. To see that, let us take a look at the spacetime decomposed Einstein-Hilbert Lagrangian \cite{Bergshoeff:2019ctr}
		\begin{eqnarray}
			\cL &=& 2 R(\Omega) \wedge T + 2 \e_{ab} R(\Omega^a) \wedge E^b \,,
			\label{Decomposed3D}
		\end{eqnarray}
		where
		\begin{align}
			R(J) &= d \Omega - \e_{ab} \Omega^a \wedge \Omega^b \,, & R^a (\Omega) & = d\Omega^a - \e^a{}_b \Omega \wedge \Omega^b \,,
		\end{align}
		where we used the fact that in two dimensions $\Omega^{ab}$ can be written as $\Omega^{ab} =  \e^{ab} \Omega$. The expansion of the gauge fields \eqref{ExpGF} gives rise to the following Lagrangian
		\begin{eqnarray}
			\cL &=& 2 \sum_{m=0}^{N_0} \sum_{n=0}^{N_0} \lambda^{2m+2n} R(\o_{(2m)}) \wedge \tau_{(2n)} + 2 \e_{ab} \sum_{m=0}^{N_1} \sum_{n=0}^{N_1} \lambda^{2m+2n+2} R(\o^a_{(2m+1)}) \wedge e^b_{(2n+1)} \,. \qquad 
		\end{eqnarray}
		This form of the Lagrangian suggests that unless $N_0 = N_1+1$ we cannot obtain all contribution to any given $\lambda^{2n}$-order. For instance, let us consider the expansion of the Lagrangian to $\lambda^2$-order
		\begin{eqnarray}
			\cL &=& \left(2 R(\o) \wedge \tau \right) + \lambda^2 \left(2 \left(R(\o) \wedge m + R(s) \wedge \tau + \e_{ab} R^a(\o) \wedge e^b \right) \right) \,,
		\end{eqnarray}
		where
		\begin{align}
			R(\o) & = d\o \,, & R(s) & = ds - \e_{ab} \o^a \wedge \o^b \,, & R^a(\o) & = d\o^a - \e^{a}{}_b \, \o \wedge \o^b \,.
			\label{Curvatures3d}
		\end{align}
		The zeroth-order action is the three-dimensional Galilei gravity. The $\lambda^2$-order theory is known as the extended Bargmann gravity \cite{Papageorgiou:2009zc}. The complete $\lambda^2$ contribution to the Lagrangian occurs as long as $N_0 =1$ and $N_1 = 0$.  Consequently, the necessary transformation rules for the gauge fields can be found by truncating the infinite-dimensional algebra \eqref{NRInfinite} by keeping  the number of even generators  two more than the number of odd generators,  namely $N_0=N_1+1$. In the case of extended Bargmann gravity, the corresponding truncation is known as the extended Bargmann algebra \cite{Papageorgiou:2009zc}
		\begin{align}
			\left[G_a, H\right] & = P_a \,,  &\left[J, P_a\right] &= -\e_{ab} P^b \,, &\left[J, G_a\right] &= -\e_{ab} G^b \nn\\
			\left[G_a, P_b \right] & = \d_{ab} M \,, & \left[G_a, G_b\right] & = - \e_{ab} S \,,
			\label{3dEBA}
		\end{align}
		where we set $J_{ab} = \e_{ab} J$ and $S_{ab} = \e_{ab} S$. In the next order for the consistent truncation, one obtains the Extended Newtonian Gravity \cite{Ozdemir:2019orp}, which we defer its details to Section \ref{D3NR}.
		
		\subsection{Ultra-Relativistic Algebras and Actions}
		\paragraph{}
		The ultra-relativistic higher-dimensional algebras are achieved with the following choice for the generators 
		\begin{align}
						V_0 & = \{J_{ab}, P_a\}\,, & V_1 & = \{H, G_a\} \,.
					\end{align}
					Thus, the Lie algebra expansion of the spacetime decomposed Poincar\'e algebra is given by
		\begin{align}
			\left[G_a^{(2m+1)}, P_b^{(2n)}\right] &= \d_{a b} H^{(2m+2n+1)}\,, &  \left[G_a^{(2m+1)}, H^{(2n+1)}\right] & = P_a^{(2m+2n+2)} \,,  \nn\\
			\left[J_{a b}^{(2m)}, P_c^{(2n)}\right] &=2 \d_{c[b} P_{a]}^{(2m + 2n)} \,, & \left[J_{a b}^{(2m)}, G_c^{(2n+1)}\right] &=2 \d_{c[b} G_{a]}^{(2m + 2n+1)}  \,, \nn\\
			\left[J_{a b}^{(2m)}, J_{c d}^{(2n)}\right] &= 4 \d_{[a[c}J_{d]b]}^{(2m+2n)}\,, & \left[G_a^{(2m+1)}, G_b^{(2n+1)}\right] & = J_{a b}^{(2m+2n+2)}\,. 
			\label{URInfinite}
		\end{align}
		The simplest truncation of this infinite-dimensional algebra is known as the Carroll algebra \cite{Duval:2014uoa, Duval:2014uva, Bergshoeff:2014jla, Duval:2014lpa, Hartong:2015xda,Bergshoeff:2017btm}
		\begin{align}
			[C_a, P_b] & = \d_{ab} H \,, &\left[J_{a b}, P_c\right] &= \d_{b c} P_a - \d_{a c} P_b\,, \nn\\
			\left[J_{a b}, C_{c}\right] &= \d_{b c} C_a - \d_{a c} C_b \,, &\left[J_{a b}, J_{c d}\right] &= 4 \d_{[a[c}J_{d]b]}\,,
			\label{CarAlb}
		\end{align}
		where we set $G_a^{(1)} = C_a$ to indicate that it is the generator of Carrollian boosts. Due to the change of the expansion character of the generators, the spacetime decomposed vielbein and the spin-connection are now given by
		\begin{align}
			T & = \sum_{n=0}^{N_1} \lambda^{2n+1} \tau_{(2n+1)} \,, & E^a & = \sum_{n=0}^{N_0}  \lambda^{2n} e^a_{(2n)} \,,\nn\\
			\Omega^{ab}  & = \sum_{n=0}^{N_0} \lambda^{2n} 	\omega^{ab}_{(2n)} \,, &  \Omega^a & = \sum_{n=0}^{N_1}  \lambda^{2n+1} \omega^a_{(2n+1)} \,.
			\label{ExpUR}
		\end{align}
		As a result, the spacetime decomposed $D=4$ Einstein-Hilbert action is expanded as
		\begin{eqnarray}
			\cL &=& -2 \e_{abc} \sum_{m=0}^{N_0} \sum_{n=0}^{N_0} \sum_{\ell=0}^{N_1} \lambda^{2m+2n+2\ell +1}  R(\omega^{ab}_{(2m)}) \wedge e^c_{(2n)} \wedge \tau_{(2\ell+1)} \nn\\
			&& + 2 \e_{abc} \sum_{m=0}^{N_1} \sum_{n=0}^{N_0} \sum_{\ell=0}^{N_0} \lambda^{2m+2n+2\ell +1} R(\omega^a_{(2m+1)}) \wedge  e^b_{(2n)} \wedge  e^c_{(2\ell)}  \,.
			\label{4dURExp}
		\end{eqnarray}
		Note that this case is much simpler than the non-relativistic case. All terms in the Lagrangian are expanded to the same $\lambda$-order. Hence, it is guaranteed to have all possible contributions to any given $\lambda$-order as long as we have equal number of even and odd generators. At the lowest order $(\l)$ we have the Carroll gravity \cite{BergshoeffCarroll,MatulichCarroll, Hartong:2015xda,Bergshoeff:2017btm,Guerrieri:2021cdz}
		\begin{eqnarray}
			\cL &=& -2 \e_{abc} R^{ab}(\o) \wedge e^a \wedge \tau + 2 \e_{abc} R^a(\o) \wedge e^b \wedge e^c \,,
		\end{eqnarray}
		where we set
		\begin{align}
			\tau_{(1)} & = \tau \,, &e^a_{(0)} &= e^a \,, & \omega^{ab}_{(0)} &= \omega^{ab} \,, & \omega^a_{(1)} & = \omega^a  \,.
			\label{CarrollExp}
		\end{align}
		For the next-to-leading order Lagrangian $(\l^3)$, we have 
		\begin{eqnarray}
			\cL &=&  -2 \e_{abc} R^{ab}(s) \wedge e^a \wedge \tau  -2 \e_{abc} R^{ab}(\o) \wedge t^a \wedge \tau -2 \e_{abc} R^{ab}(\omega) \wedge e^a \wedge m \nn\\
			&& + 2 \e_{abc} R^a(b) \wedge e^b \wedge e^c + 4 \e_{abc} R^a(\o) \wedge e^b \wedge t^c  \,.
		\end{eqnarray}
		where we have 
		\begin{align}
			\tau_{(1)} & = \tau \,, & \tau_{(3)} & = m \,, & \omega^{ab}_{(0)} &= \omega^{ab} \,, & \omega^{ab}_{(2)} & = s^{ab} \,,\nn\\
			e^a_{(0)} &= e^a \,, & e^a_{(2)} &= t^a \,, &  \omega^a_{(1)} & = \omega^a &  \omega^a_{(3)} & = b^a \,. 
			\label{BeyondCarrollExp}
		\end{align}
		Here, the group theoretical curvatures read
		\begin{align}
			R^{ab}(\o) & = d\o^{ab} - \o^{ac} \wedge \o^b{}_c \,, & R^a(\o) & = d\o^a + \o^{ab} \wedge \o_b \,, \nn\\
			R^{ab}(s) & = ds^{ab} - 2 \o^{ac} \wedge s^b{}_c - \o^a \wedge \o^b \,, & R^a(b) & = db^a + \o^{ab} \wedge b_b + s^{ab} \wedge \o_b \,.
		\end{align}
		This Lagrangian is invariant under the following extension of the Carroll algebra
		\begin{align}
			[C_a, P_b] & = \d_{ab} H \,, &\left[J_{a b}, P_c\right] &= \d_{b c} P_a - \d_{a c} P_b\,, & 	\left[J_{a b}, C_{c}\right] &= \d_{b c} C_a - \d_{a c} C_b \,, \nn\\
			\left[J_{a b}, J_{c d}\right] &= 4 \d_{[a[c}J_{d]b]}\,, & 	[B_a, P_b] & = \d_{ab} M \,, & [C_a, T_b] & = \d_{ab} M \,, \nn\\
			[C_a, H] & =T_a \,, & \left[J_{a b}, T_c\right] &= \d_{b c} T_a - \d_{a c} T_b\,, & \left[S_{a b}, P_c\right] &= \d_{b c} T_a - \d_{a c} T_b\,,  \nn\\ 
			\left[J_{a b}, B_c\right] &= \d_{b c} B_a - \d_{a c} B_b\,, & \left[S_{a b}, C_c\right] &= \d_{b c} B_a - \d_{a c} B_b\,,  & \left[J_{a b}, S_{c d}\right] &= 4 \d_{[a[c}S_{d]b]}\,,\nn\\
			\left[G_a, G_b\right] & = S_{a b}\,,
			\label{ExtCarAlb}
		\end{align}
 		where we labeled $H^{(3)} = M$ and $P_a^{(2)} = T_a$.  To our knowledge, it is still an open problem to show the equivalence of the first-order action \eqref{BeyondCarrollExp} and the second order beyond-Carrollian action in \cite{Hansen:2021fxi}. Nevertheless, we refer to this model as the beyond-Carrollian gravity. Note that 
 		all terms in the spacetime decomposed three-dimensional Einstein-Hilbert Lagrangian are still in the same order in the expansion, see \eqref{Decomposed3D}. Thus, the argument that we present for consistency of truncation also applies to three dimensions and we will not present $D=3$ as an exceptional case.

		\section{Non-Relativistic Scaling Limit of Multimetric Gravity}\label{MultiGrav}
		\paragraph{}
		
		In this section, we introduce a non/ultra-relativistic scaling limit for multimetric gravity models based on the Lie algebra expansion that we discussed in the previous section. As mentioned, the fundamental idea to have a well-defined scaling limit is the necessity of using multiple copies of Einstein-Hilbert action to get rid of divergent lower order Lagrangian(s). To provide a concrete example, let us consider the non-relativistic expansion of Einstein-Hilbert action to order $\lambda$, i.e.
		\begin{eqnarray}
			\cL &=& M_1^2 \lambda \left(-2 \e_{abc} R^{ab}(\o) \wedge e^a \wedge \tau \right)  + \mathcal{O}(\l^3)
		\end{eqnarray}
		where $M_1$ is a mass parameter that multiplies the Einstein-Hilbert action. Note that we expand the fields based on their expansion character \eqref{ExpGF} 
		\begin{align}
			\tau_{(0)} & = \tau \,, & e^a_{(1)} & =  e^a \,, & \omega^{ab}_{(0)} & = \omega^{ab} \,, & \omega^a_{(1)} =  \omega^a \,.
		\end{align}
		If we rescale the mass parameter $M_1^2$ as $M_1^2 \to M_1^2 / \lambda$ and take the limit $\lambda \to 0$, we precisely recover Galilei gravity as a non-relativistic limit of General Relativity. Let us now consider the next order in expansion
		\begin{eqnarray}
			\cL &=&M_1^2 \lambda \left(-2 \e_{abc} R^{ab}(\o) \wedge e^c \wedge \tau \right) \nn\\
			&& + M_1^2 \lambda^3 \Big( -2 \e_{abc} \left(R^{ab}(\o) \wedge e^c \wedge m + R^{ab}(s) \wedge e^c \wedge \tau + R^{ab}(\o) \wedge t^c \wedge \tau  \right)\nn\\
			&& \qquad \qquad  + 2 \e_{abc} R^a(\o) \wedge e^b \wedge e^c  \Big) + \mathcal{O}(\lambda^5) \,,
		\end{eqnarray} 
		where the fields are expanded in accordance with \eqref{NewtonExpand}. Clearly, to single out $\lambda^3$ action, it is not sufficient to rescale the mass parameter $M_1^2 \to M_1^2/\lambda^3$ and take the $\lambda \to 0$ limit since the coefficient of the Galilei gravity diverges in that limit. Thus, we must first annihilate the Galilei gravity action, then perform the proper scaling of the mass parameter and the limit. This can be achieved by considering two copies of the Einstein-Hilbert action
		\begin{eqnarray}
			\cL &=& M_1^2  \left[2\epsilon_{a b c} \left(R(\Omega^{a})\wedge E^b \wedge E^c - R(\Omega^{a b})\wedge E^c \wedge T \right)\right] \nn\\
			&& + M_2^2  \left[2\epsilon_{a b c} \left(R(\bar \Omega^{a})\wedge \bar E^b \wedge \bar E^c - R(\bar\Omega^{a b})\wedge \bar E^c \wedge \bar T \right)\right] \,,
		\end{eqnarray}
		where the second set of gauge fields are $\{\bar T\,, \bar E^a\,, \bar \O^a\,, \bar \O^{ab} \}$. To cancel out the Galilei gravity, we may keep the definitions of the expansion of the first set of fields as in \eqref{NewtonExpand} but expand the second set of the fields as \eqref{ExpGF} with the following definitions to cancel out lowest order divergences as long as $M_1^2 = M_2^2$
		\begin{align}
			\bar	\tau_{(0)} & = - \tau \,, & \bar \tau_{(2)} & = m \,, &\bar  \omega^{ab}_{(0)} &= \omega^{ab} \,, &\bar  \omega^{ab}_{(2)} & =-  s^{ab} \,,\nn\\
			\bar	e^a_{(1)} &= e^a \,, & \bar e^a_{(3)} &= - t^a \,, &  \bar \omega^a_{(1)} & = \omega^a & \bar  \omega^a_{(3)} & = b^a \,. 
			\label{NewtonExpandC2}
		\end{align}
		With this choice for fields, the Galilei gravity comes with an opposite sign and cancels the contribution from the first copy of the Einstein-Hilbert action. On the other hand, the $\mathcal{O}(\l^3)$ terms come with the same numerical factors and signatures. Consequently, if we incorporate the following scaling limit for the mass parameters
		\begin{align}
			M_1^2 = M_2^2  =  \frac{M^2}{2\lambda^3}  \,,
		\end{align}
		we precisely recover the first-order formulation of Newtonian gravity after taking the limit $\lambda \to 0$. In this section, our purpose is to define the systematic of non/ultra-relativistic scaling limit of multimetric gravity in this line of consideration. Before proceeding to the actual computation, we remind the reader that the invariant non-relativistic actions in dimensions $D\geq 4$ require an equal number of even and odd fields in the expansion while $D=3$ is an exception, requiring two more even fields than the odd fields \cite{Bergshoeff:2019ctr}. Thus, we shall investigate these two cases separately. 
		
		The spacetime decomposed Einstein-Hilbert action in $D$-dimensions take the following form
		\begin{eqnarray}
			\cL_{EH} &=& 2 \e_{a_1 a_2 \ldots a_{d-1} a_d} \,  E^{a_1} \wedge \ldots \wedge E^{a_{d-1}} \wedge R(\Omega^{a_d}) \nn\\
			&& + (d-1)  \e_{a_1 a_2 \ldots a_{d-1} a_d} \, T \wedge E^{a_1} \wedge \ldots \wedge E^{a_{d-2}} \wedge R(\Omega^{a_{d-1} a_d}) \,,
			\label{DExp}
		\end{eqnarray}
		where $d = D-1$ is the number of spatial dimensions. For the non-relativistic expansion, the form of the relativistic fields \eqref{ExpGF} suggests that we should investigate $d=2$ and $d > 2$ cases separately since for $d=2$ all $E^a$ terms drop out in the second term of the Lagrangian. The reason for this is the following. Consider the expansion of the fields to order $(N_0, N_1)$ where we either have $N_0 = N_1$ or $N_0 = N_1 +1$. It is sufficient to consider the appearance of the highest order even and odd fields to see which choice include all terms at a given order $\mathcal{O}(\lambda^n)$.  If the highest odd field is of order $\l^{2N+1}$, then we have two choices for the highest even field depending on the chosen consistent truncation condition:
		\begin{itemize}
			\item {The highest even field is $\mathcal{O}(\lambda^{2N})$: In this case, the second term in the Lagrangian \eqref{DExp} indicates that both the highest even and the odd fields appears at order $d - 2 + 2N $.}
			\item{The highest even field is $\mathcal{O}(\lambda^{2N+2})$: In this case, the second term in the Lagrangian \eqref{DExp} indicates that the highest odd field appears at order $d - 2 + 2N$ while the highest order even field appear at order $d + 2N$. This case cannot define a complete invariant Lagrangian since we terminate the odd fields at $\mathcal{O}(\lambda^{2N+1})$, leading us to miss the contributions of $\mathcal{O}(\lambda^{2N+3})$ odd fields to the action of order $d+2N$.}
		\end{itemize}
		Note that the situation changes dramatically for $d=2$. In that case, the second term in \eqref{DExp} involves no odd field, and, as explained in Section \ref{NRPrelim}, the consistent truncation occurs by setting $N_0 = N_1 + 1$. Thus, in this section, we shall separately investigate $d=2$ and $d > 2$. For $d > 2$, we choose $D=4$ as a representative example, however our arguments can be straightforwardly generalized to $D > 4$.
		
		\subsection{\texorpdfstring{$D=4$}{}}\label{NR4}
		\paragraph{}
		In $D=4$, the spacetime decomposed Einstein-Hilbert takes the following form 
		\begin{eqnarray}
			\cL_{EH} &=& 	2M^2 \epsilon_{a b c} \left(R(\Omega^{a})\wedge E^b \wedge E^c - R(\Omega^{a b})\wedge E^c \wedge T \right) \,.
	\label{4dEH}
	\end{eqnarray}
		As mentioned, the Lie algebra expansion of this Lagrangian is formally given by
		\begin{eqnarray}
			\cL_{\rm EH} &=& \l M^2 \cL_{1} + \l^3 M^2 \cL_{3} + \l^5 M^2 \cL_{5} + \ldots \,.
			\label{NREH}
		\end{eqnarray}
		where $\cL_{n}$ represents the Lagrangian at the $\lambda^n$-order.  This form of the Lagrangian implies that if we want to single out a $\mathcal{O}(\l^N)$ Lagrangian by eliminating all terms of order $n < N$, rescaling the mass parameter by $\lambda^{-N}$ and performing the $\lambda \to 0$ scaling limit, we need $N$-copies of the Einstein-Hilbert action. An example of non-interacting bi-metric gravity and its limit as Newtonian gravity was introduced in Section \ref{MultiGrav} by keeping the same expansion for the first set of gauge fields \eqref{ExpGF} and properly choosing the second set to annihilate the $\mathcal{O}(\l)$ action. For a more general result, consider a non-interacting model of multimetric  gravity that involves $N$-number of spacetime decomposed vielbein and spin-connection
		\begin{eqnarray}
			\cL &=& \sum_{i=1}^N \left[2M_{(i)}^2 \epsilon_{a b c} \left(R(\Omega^{a}_{(i)})\wedge E^b_{(i)} \wedge E^c_{(i)} - R(\Omega^{a b}_{(i)})\wedge E_{(i)}^c \wedge T_{(i)} \right)  \right] \,,
		\end{eqnarray}
		where $i,j=1,\ldots N$ label the set of fields $\{T_{(i)},E_{(i)}, \Omega^{ab}_{(i)}, \Omega^a_{(i)}\}$. Following our example for the Newtonian gravity, we keep the expansion of the first set of fields $\{T_{(1)}, E_{(1)}, \Omega^a_{(1)}, \Omega^{ab}_{(1)}\}$ the same as \eqref{ExpGF}, that is,
		\begin{align}
			T_{(1)} & = \sum_{n=0}^{N_0} \lambda^{2n} \tau_{(2n)} \,, & E^a_{(1)} & = \sum_{n=0}^{N_1}  \lambda^{2n+1} e^a_{(2n+1)} \,,\nn\\
			\Omega^{ab}_{(1)}  & = \sum_{n=0}^{N_0} \lambda^{2n} 	\omega^{ab}_{(2n)} \,, &  \Omega^a_{(1)} & = \sum_{n=0}^{N_1}  \lambda^{2n+1} \omega^a_{(2n+1)} \,.
			\label{ExpGF1}
		\end{align}
		However, for $i > 1$, we can incorporate dimensionless free parameters $\alpha_{(i)}$ as follows\footnote{ The expansion that we discuss here include the previous example \eqref{NewtonExpandC2} as a special case up to an overall scaling and the change of signature of the fields.}
		\begin{align}
			T_{(i)} & = \sum_{n=0}^{N_0} \lambda^{2n} \alpha_{(i)}^{2n} \tau_{(2n)} \,, & E^a_{(i)} & = \sum_{n=0}^{N_1}  \lambda^{2n+1} \alpha_{(i)}^{2n+1}  e^a_{(2n+1)} \,,\nn\\
			\Omega^{ab}_{(i)}  & = \sum_{n=0}^{N_0} \lambda^{2n} \alpha_{(i)}^{2n} 	\omega^{ab}_{(2n)} \,, &  \Omega^a_{(i)} & = \sum_{n=0}^{N_1}  \lambda^{2n+1} \alpha_{(i)}^{2n+1}  \omega^a_{(2n+1)} \,.
			\label{AlphaExpansion}
		\end{align}
where $\alpha_i \neq \alpha_j$ for $i \neq j$ and $\alpha_i \neq 1$ which are the necessary conditions for having linearly independent set of gauge fields. This expansion would yield the following expansion of $\cL_{(i)}$ that is the Einstein-Hilbert action for $i$-th set of fields $\{T_{(i)}, E_{(i)}, \Omega^a_{(i)}, \Omega^{ab}_{(i)}\}$ 
		\begin{eqnarray}
			\cL_{EH,(i)} &=&  \l \alpha_{i} M_{i}^2 \cL_{1} + \l^3 \alpha_{i}^3 M_{i}^2 \cL_{3} + \l^5 \alpha_{i}^5 M_{i}^2 \cL_{5} + \ldots \,.
		\end{eqnarray}
		Note that the structure of the Lagrangian at each order $\lambda$ does not change by the inclusion of the free parameters but we simply pick up a free coefficient $\alpha_{i}^{2n+1}$ in front of the Lagrangians that are necessary for cancellations. Let us now work out certain examples and then provide a general formalism.
		\begin{enumerate}
			\item {\textbf{Galilei Gravity:} The Galilei gravity appears as $\cL_1$ in the expansion of Einstein-Hilbert action. As mentioned, it is sufficient to  consider a single copy of Einstein-Hilbert action and express the relativistic fields and the mass parameter as
				\begin{align}
					T &= \tau \, & E^a &=\lambda e^a \,,&  \Omega^{ab} & = \omega^{ab} \,, & \Omega^a & = \lambda \omega^a \,, & M^2 & = \frac{m^2}{\lambda} \,.
				\end{align}
				Upon taking the $\lambda \to 0$ limit, this choice would lead to the Galilei gravity.}
			\item{\textbf{Newtonian Gravity:} The Newtonian gravity appears at the $\lambda^3$-order in the Lie algebra expansion. Thus, we shall consider two copies of Einstein-Hilbert action. For the first copy, we keep the same form as the Lie algebra expansion, i.e.
				\begin{align}
					T_{1}  &= \tau + \lambda^2 m \,, & E^a_{1}  &= \lambda e^a + \lambda^3 t^a \,,&	\Omega^{ab}_{1}  & = \omega^{ab} + \l^2 s^{ab} \,, & \Omega^a_{1}  & = \lambda \omega^a + \lambda^3 b^a \,.
				\end{align}
				For the second copy, we incorporate the free parameter $\alpha_2$ 
				\begin{align}
					T_{2}  &= \tau + \lambda^2 \alpha_2^2 m \,, & E^a_{2}  &= \lambda \alpha_2 e^a + \lambda^3 \alpha_2^3 t^a \,,\nn\\
					\Omega^{ab}_{2}  & = \omega^{ab} + \l^2 \alpha_2^2 s^{ab} \,, & \Omega^a_{2}  & = \lambda \alpha_2 \omega^a + \lambda^3 \alpha_2^3 b^a \,.
				\end{align}
				The combination of these two Lagrangian with mass parameters $M_{1,2}$ gives rise to
				\begin{eqnarray}
					\cL &=& \lambda M^2 \left(1 + \frac12 \alpha_2\right) \cL_1 + \lambda^3 M^2 \left(1+ \frac12 \alpha_2^3\right) \cL_3 + \mathcal{O}(\lambda^5) \,.
				\end{eqnarray}
				where we set $M_1^2 = 2M_2^2 = M^2$ for simplicity. To cancel out the first term, we fix $\alpha_2$ to be $\alpha_2 = - 2$. The coefficient of $\cL_3$ then becomes $-3$.  Then, the following scaling of the mass parameter
				\begin{eqnarray}
					M^2 = \frac{1}{3 \lambda^3} m^2 \,.
				\end{eqnarray}
				along with the scaling limit $\lambda \to 0$ recover the action principle for the Newtonian gravity up to an overall minus sign.
			}
			\item{\textbf{Beyond the Newtonian Gravity:} The next-to-Newtonian gravity action arise at order $\mathcal{O}(\lambda^5)$. To perform a proper scaling limit, we need three copies of the Einstein-Hilbert action. Using the expansion of the relativistic fields \eqref{AlphaExpansion}, we obtain the following Lagrangian
				\begin{eqnarray}
					\cL &=&   \lambda \left(M_1^2 + M_2^2 \alpha_2 + M_3^2 \alpha_3 \right) \cL_1 + \lambda^3 \left(M_1^2 + M_2 ^2\alpha_2^3 + M_3 ^2\alpha_3^3\right) \cL_3  \nn\\
					&& + \lambda^5 \left(M_1^2 + M_2 ^2\alpha_2^5 + M_3 ^2\alpha_3^5 \right) \cL_5 +  \mathcal{O}(\lambda^7)  \,.
				\end{eqnarray}
				Here, $\cL_5$ represents the next-to-Newtonian gravity Lagrangian.  For convenience, let us choose $M_1^2 = 2 M_2^2 = 2/3 M_3^2 = M^2$. In that case, to cancel out the coefficients of the Galilei and Newtonian gravity, we need to solve two equations
				\begin{align}
					0 & = 1 + \frac12 \alpha_2 + \frac32 \alpha_3  \,, & 0 & = 1 + \frac12 \alpha_2^3 +  \frac32 \alpha_3^3 \,.
				\end{align}
				These two equations can be solved to eliminate $\alpha_2$ and $\alpha_3$ as $\alpha_2 = -5/4$ and $\alpha_3 = -1/4$. Then, the following scaling of the mass parameter
				\begin{eqnarray}
					M^2 = \frac{256}{135 \lambda^5} m^2 \,.
				\end{eqnarray}
				along with the scaling limit $\lambda \to 0$ recover the Lagrangian for next-to-Newtonian gravity up to an overall minus sign.
			}
		\end{enumerate}
		With these three examples, it is now evident that for a Lagrangian of order-$2N+1$, we need to introduce $N$-copies of Einstein-Hilbert action. Using the expansion of the fields, we obtain 
		\begin{eqnarray}
			\cL &=& \sum_{n=0}^{N} \l^{2n+1} \left( M_1^2 + \sum_{i=2}^{N+1} \a_i^{2n+1} M_i^2 \right) \cL_{2n+1} + \mathcal{O}(\l^{2N+3}) \,.
		\end{eqnarray}
		To single out the order-$2N+1$ Lagrangian, one needs to solve $N$ number of algebraic equations that determines the values of $\alpha_i$ for $i = 2, 3, \ldots N$
		\begin{eqnarray}
			0&=&	 M_1^2 + \sum_{i=2}^{N+1} \a_i^{2n+1} M_i^2 \,, \qquad \text{for} \qquad n = 0,1,\ldots, N-1 \,.
		\end{eqnarray}
		These values can finally be used in the coefficient of the order-$2N+1$ Lagrangian, along with the scaling of the mass parameters $M_i \to M_i/\lambda^{2N+1}$. Finally, performing the $\lambda \to 0$ limit gets rid of all higher order terms and yields the desired non-relativistic model.
		
		Although we have so far discussed the non-relativistic limit of non-interacting multimetric models, we can turn on potential terms that gives rise to the interaction among $E_{(i)}^a$. Upon spacetime decomposition, they typically take the following form in four dimensions
		\begin{eqnarray}
			\cL_{CC} = \e_{abc} J_{ijkl} T_{i} \wedge E^a_j \wedge E^b_k \wedge E^c_l \,,
			\label{CosmologicalConstant}
		\end{eqnarray}
		where $J_{ijjk}$ is a matrix of constant coefficients of dimension mass-squared. The off-diagonal elements of this matrix determine the interaction between the temporal and spatial vielbein of different gravity sectors. Upon implementing the $\alpha$-expansion, \eqref{ExpGF1} and \eqref{AlphaExpansion}, we notice that the lowest order contribution arise at $\mathcal{O}(\lambda^3)$, i.e.
		\begin{eqnarray}
			\cL &=& \lambda^3 \e_{abc} \tau \wedge e^a \wedge e^b \wedge e^c + \mathcal{O}(\l^5) \,,
		\end{eqnarray}
		where we assumed that the coefficients of $J_{ijkl}$ and $\alpha_{(i)}$ are not chosen in a particular fashion to annihilate each other. This is the same expansion order as the Newtonian gravity. As we will discuss the details in Appendix \ref{AppA}, this expansion implies that the non-relativistic limit of bi-gravity with a non-vanishing potential is the Newtonian gravity with a constant background mass density. For the three-metric model, we need to go to the $\mathcal{O}(\l^5)$
		\begin{eqnarray}
			\cL \sim   \l^5 \e_{abc} \left( m \wedge e^a \wedge e^b \wedge e^c  + 3 \tau \wedge t^a\wedge e^b\wedge  e^c\right) + \mathcal{O}(\l^7) \,. 
		\end{eqnarray}
		Note that we have more than a sufficient number of free parameters to annihilate the order-$\lambda^3$ Lagrangian, hence the details are not presented.
		\subsection{\texorpdfstring{$D=3$}{}}\label{D3NR}
		\paragraph{}
		In three-dimensions, the spacetime decomposed Einstein-Hilbert Lagrangian takes the following form \eqref{Decomposed3D} 
		\begin{eqnarray}
			\cL &=&  2 M  \left( R(\Omega) \wedge T +  \e_{ab} R(\Omega^a) \wedge E^b \right)  \,,
		\end{eqnarray}
		which implies that upon Lie algebra expansion \eqref{ExpGF} we have the following structure for the Lagrangian
		\begin{eqnarray}
			\cL_{EH} &=& \l^0 M \cL_0 + \l^2 M \cL_2 + \l^4 M \cL_4 + \ldots \,.
		\end{eqnarray}
		Let us start our investigation with the zeroth-order Lagrangian which describes the three-dimensional Galilei gravity
		\begin{eqnarray}
			\cL_0 &=& 2 \tau \wedge d\omega \,.
		\end{eqnarray}
		This model only involves the gauge fields of time translation and rotations, which, in the case of three-dimensions, is a $U(1) \times U(1)$ gauge theory. Therefore, this model can be considered as an off-diagonal $U(1) \times U(1)$ Chern-Simons gauge theory
		\begin{eqnarray}
			\cL &=&  2 Z_1 \wedge d Z_2 \,,
		\end{eqnarray}
		which the identification $Z_1 = \tau$ and $Z_2 = \omega$, corresponding to the off-diagonal invariant metric with $g(Z_1,Z_2) = 1$. At this stage, it is obvious that the elimination of lower-order Lagrangians to single out higher-order is more subtle in $D=3$. It is not sufficient to consider multiple copies of Einstein-Hilbert action, but the theory must be extended with a $U(1) \times U(1)$ Chern-Simons gauge theory to cancel out the lowest order Lagrangian, see \cite{Bergshoeff:2016lwr,Ozdemir:2019orp,Ozdemir:2019tby} as particular examples.
		
		Let's now move on to the next order $\cL_2$. In this case, we have
		\begin{eqnarray}
			\cL &=& M \left(2 R(\o) \wedge \tau \right) + \lambda^2 M \left(2 \left(R(\o) \wedge m + R(s) \wedge \tau + \e_{ab} R^a(\o) \wedge e^b \right) \right) + \mathcal{O}(\l^4) \,.
		\end{eqnarray}
		where the curvatures are as defined in \eqref{Curvatures3d}. To obtain this model with as a scaling limit, we begin with the three-dimensional Einstein-Hilbert action with an additional $U(1) \times U(1)$ Chern-Simons gauge theory
		\begin{eqnarray}
			\cL &=& 2M R(\Omega) \wedge T + 2M  \e_{ab} R(\Omega^a) \wedge E^b  + 2 M Z_1 \wedge d Z_2 \,. 
		\end{eqnarray}
		The zeroth-order Galilei gravity can then be annihilated by a proper choice of $Z_1$ and $Z_2$, i.e.
		\begin{align}
			T & = \tau + \lambda^2 m \,, & E^a & = \lambda e^a \,, & \Omega & = \omega + \lambda^2 s  \,, & \Omega^a & = \lambda \omega^a \,,\nn\\
			Z_1 & = - \tau\,, & Z_2 & = \omega \,,
		\end{align}
		which precisely recovers the $\mathcal{O}(\l^2)$ Lagrangian, known as the extended Bargmann gravity \cite{Papageorgiou:2009zc} upon rescaling the mass parameter $M \to M/\lambda^2$ and taking the limit $\lambda \to 0$. Note that this contraction is much simpler than the one presented in \cite{Bergshoeff:2016lwr}, thank to the Lie algebra expansion. Similarly, if we are to move on with the next order Lagrangian, we have
		\begin{eqnarray}
			\cL &=& M \left(2 R(\o) \wedge \tau \right) + \lambda^2 M \left(2 R(\o) \wedge m + 2 R(s) \wedge \tau - 2  R^a(\o) \wedge e_a \right) \nn\\
			&& + \l^4 M \left( R(s) \wedge m + R(z) \wedge \tau - R(\omega) \wedge y- R^a(\omega) \wedge t_a - R^a(b) \wedge e_a \right)  + \mathcal{O}(\l^6) 
		\end{eqnarray}
		where the $\mathcal{O}(\l^4)$ Lagrangian is known as the extended Newtonian gravity \cite{Ozdemir:2019orp}. Note that the Lagrangian involves the redefinition of the fields 
		\begin{align}
			e^a & \to \e^{ab} e_b\,, & t^a & \to \e^{ab} t_b \,, & y \to - y \,,
		\end{align}
		in the Lie algebra expansion of the fields to match with the existing literature \cite{Bergshoeff:2019ctr}. As in the case of four-dimensions, we can reproduce this Lagrangian by considering a three-dimensional bigravity with an additional $U(1) \times U(1)$ Chern-Simons gauge theory
		\begin{eqnarray}
			\cL &=& 2 M_1 R(\Omega_1) \wedge T_1 + 2 M_1 \e_{ab} R(\Omega^a_1) \wedge E^b_1 + 2 M_2 R(\Omega_2) \wedge T_2 \nn\\
			&& + 2 M_2 \e_{ab} R(\Omega^a_2) \wedge E^b_2 + 2 M_3 Z_1 \wedge d Z_2 \,.
		\end{eqnarray}
		Using the standard expansion for the first set of fields and  implementing the $\alpha$-expansion in the second set of fields
		\begin{align}
			T_1 & = \tau + \lambda^2 m - \lambda^4 y \,, &\Omega_1 & = \omega + \lambda^2 s + \lambda^4 z \,, & \Omega^a_1 & = \lambda \o^a + \lambda^3 b^a \,,\nn\\
			E^a_1 & = \l \e^{ab} e_b + \l^3 \e^{ab} t_b \,, & Z_1 & = \beta_1 \tau \,, & Z_2 & = \omega \,,\nn\\
			T_2 & = \tau + \alpha_2^2 \lambda^2 m - \alpha_2^4 \lambda^4 y \,, &\Omega_2 & = \omega +\alpha_2^2 \lambda^2 s +\alpha_2^4 \lambda^4 z \,, & \Omega^a_2 & = \alpha_2 \lambda \o^a +\alpha_2^3 \lambda^3 b^a \,,\nn\\
			E^a_2 & = \alpha_2 \lambda \e^{ab} e_b +\alpha_2^3  \l^3 \e^{ab} t_b \,, 
			\label{NewtonianScaling}
		\end{align}
		we obtain
		\begin{eqnarray}
			\cL &=& \left(M_1 + \beta_1 M_3 +  M_2 \right) \left(2 R(\o) \wedge \tau \right) \nn\\
			&& + \l^2  \left(M_1  + \alpha_2^2 M_2 \right)  \left(2 R(\o) \wedge m + 2 R(s) \wedge \tau - 2  R^a(\o) \wedge e_a \right)  \nn\\
			&& +  \l^4 \left(M_1  + \alpha_2^4 M_2 \right)  \left( R(s) \wedge m + R(z) \wedge \tau - R(\omega) \wedge y- R^a(\omega) \wedge t_a - R^a(b) \wedge e_a \right)  \nn\\
			&& + \mathcal{O}(\l^6) \,.
		\end{eqnarray}
		Note that the three-dimensional case is not complicated than the scaling limit in four-dimensions. The $U(1) \times U(1)$ gauge theory only annihilates the zeroth-order Lagrangian but does not interfere with higher-order ones. Consequently, $\beta_1$ can be solved by using the coefficient of the Galilei gravity and the remainder of the problem is the same as the four-dimensional case. To present a solution, let us set $M_1 = M_3 = - 4 M_2 = M$. In that case, we can set the coefficients of the Galilei and the extended Bargmann gravity by the following choices for $\alpha_2$ and $\beta_1$
		\begin{align}
			\alpha_2 &= 2 \,, & \beta_1 & = - \frac{3}{4} \,.
		\end{align}
		Finally, rescaling the mass parameter $M$ as
		\begin{eqnarray}
			M = - \frac{m}{3 \lambda^4} \,,
		\end{eqnarray}
		and performing the scaling limit $\lambda \to 0$ precisely recover the extended Newtonian gravity \cite{Ozdemir:2019orp}. Once again, the scaling limit presented in \eqref{NewtonianScaling} is much simpler than the one found in  \cite{Ozdemir:2019orp}.
		
		With these three examples, the systematic of three-dimensional non-relativistic scaling limit is obvious. To start with , one needs an $N$-copy of Einstein-Hilbert action with an additional $U(1) \times U(1)$ Chern-Simons gauge theory
		\begin{eqnarray}
			\cL &=& \sum_{i=1}^N \left[ 2M_{(i)} R(\Omega_{(i)}) \wedge T_{(i)} + 2M_{(i)}  \e_{ab} R(\Omega^a_{(i)}) \wedge E^b_{(i)} \right] + 2 M_{N+1} Z_1 \wedge d Z_2 \,.
		\end{eqnarray}
		The fields can then be expanded using the standard expansion for the first set of fields \eqref{ExpGF} and the $\alpha$-expansion for the other set of fields \eqref{AlphaExpansion}. In addition, the gauge fields $Z_1$ and $Z_2$ must be chosen as
		\begin{align}
			Z_1 =  &\beta_1 \tau \,, &   Z_2 & = \omega \,.
		\end{align}
		These choices would lead us to the following Lagrangian for the $N$-metric theory
		\begin{eqnarray}
			\cL &=& \left( \beta_1 M_{N+1} +   \sum_{i=1}^N M_{i}  \right) \cL_0 + \sum_{n=1}^{N} \l^{2n} \left( M_1 + \sum_{i=2}^{N} \a_i^{2n+1} M_i \right) \cL_{2n} + \mathcal{O}(\l^{2N+2}) \,.
		\end{eqnarray}
		The solution for $\beta_1$ is simply
		\begin{eqnarray}
			\beta_1 =  - \frac{1}{M_{N+1}} \sum_{i=1}^N M_{(i)}  
		\end{eqnarray}
		For the remaining coefficients, we can follow our footsteps in $D=4$ and solve the coefficients of the $\mathcal{O}(\lambda^{2N-2})$ models to single out the Lagrangian at order $\lambda^{2N}$. In the last step, we rescale the mass parameters $M_{(i)} \to M_{(i)} / \lambda^{2N}$ and perform the scaling limit $\lambda \to 0$ and obtain the desired non-relativistic Lagrangian.
		
		\subsection{Non-Relativistic Algebras as a Contraction of Multiple Poincar\'e Algebras}
		\paragraph{}
		The Lie algebra expansion is not just the expansion of the gauge fields but also the expansion of the generators \eqref{GeneratorExpansion}. Thus, it is expected that the scaling limit that we establish here for the Lagrangians can be extended to a relation between $N$-copies of the Poincar\'e algebra and the non-relativistic algebra at the relevant order. The Galilei and the Bargmann algebra have been known to arise from the contraction of the Poincar\'e and Poincar\'e $\oplus $ U(1) algebras, respectively \cite{Andringa:2010it}. In this section, we show that the higher-order non-relativistic algebras that admit an invariant Lagrangian arise from the contraction of multiple copies of Poincar\'e algebra. There is of course an exception in three-dimension with requires an addition of $U(1) \times U(1)$ sector to cancel out divergences.
		
		Let us start our discussion with two representative examples: The Galilei algebra \eqref{GalAlb} as a contraction of the Poincar\'e algebra and the non-relativistic algebra of Newtonian gravity \eqref{NewtAlg} as a contraction of two copies of the Poincar\'e algebra. Consider the spacetime decomposed Poincar\'e algebra \eqref{decomposealgebra}. If we rescale the odd generators as
		\begin{align}
			P_a & \to \lambda P_a \,, & G_a \to \lambda G_a \,,
		\end{align}
		and take the $\lambda \to \infty$ limit, the $[P_a, G_b]$ and $[G_a, G_b]$ commutators vanish, giving rise to the Galilei algebra \eqref{GalAlb}. Next, consider two copies of the spacetime decomposed Poincar\'e algebra, the first set being labeled as $\{H^1, P_a^{1}, G_a^{1}, J_{ab}^{1}\}$ and the second set with $\{H^{2}, P_a^{2}, G_a^{2}, J_{ab}^2 \}$. Expressing the non-relativistic generators as
		\begin{align}
			J_{ab} & =  J_{ab}^{1} +  J_{ab}^{2}\,, & S_{ab} &= \l^2 \left( \alpha_1^2 J_{ab}^{1} +  \alpha_2^2 J_{ab}^{2}\right) \,,\nn\\
			H &= H^{1} + H^{2} \,, & M & = \lambda^2 \left( \a_1^2 H^{1} + \a_2^2 H^{2} \right) \,,\nn\\
			P_a & = \lambda \left( \a_1 P_a^{1} + \a_2 P_a^{2}  \right) \,, & T_a & = \lambda^3 \left( \a_1^3 P_a^{1} + \a_2^3 P_a^{2}  \right) \,,\nn\\
			G_a & = \lambda \left( \a_1 G_a^{1} + \a_2 G_a^{2}  \right) \,, & B_a & = \lambda^3 \left( \a_1^3 G_a^{1} + \a_2^3 G_a^{2}  \right) \,,
		\end{align}
		we precisely recover the commutation relations for the algebra \eqref{NewtAlg} in the limit $\l \to 0$ as long as $\a_{1}, \a_2 \neq 1$ and $\alpha_1 \neq \alpha_2$. Note that we perform the $\l \to 0$ limit rather than $\l \to \infty$ since the relations are inverted. We can use these inverted relations to establish the generators of the Poincar\'e algebra in terms of the generators of the algebra \eqref{NewtAlg}, in which case $\l \to \infty$ limit must be taken. However this relation is much easier to see that the necessary commutation relations are satisfied. As a matter of fact, the $N$-th order non-relativistic algebra \eqref{NRInfinite} can be established by using the $N$ copies of the Poincar\'e algebra
		\begin{align}
			J_{ab}^{(2n)} & = \lambda^{2n}  \bigoplus\limits_{i=1}^{N_0} \alpha_i^{2n} J_{ab}^{i} \,, & 	H^{(2n)} & = \lambda^{2n}  \bigoplus\limits_{i=1}^{N_0} \alpha_i^{2n} H^{i} \,, \nn\\
			P_a^{2n+1} &  = \lambda^{2n+1}  \bigoplus\limits_{i=1}^{N_1} \alpha_i^{2n+1} P_{a}^{i} \,, & 	G_a^{2n+1} &  = \lambda^{2n+1}  \bigoplus\limits_{i=1}^{N_1} \alpha_i^{2n+1} G_{a}^{i} \,, 
		\end{align}
		once the $\lambda \to 0$ limit is taken, given that $N_0 = N_1$. Note that this expansion makes sense as long as we impose $\alpha_i \neq 1$ and $\alpha_i \neq \alpha_j$ for $i \neq j$. In this generator expansion these alpha parameters are free and not necessary to be fixed.  Once again, these relations can be inverted  to construct the elements of the Poincar\'e algebras in terms of the generators of the non-relativistic algebra. However, with this form of the relation, it is trivial that the commutation relations \eqref{NRInfinite} are satisfied in the $\lambda \to 0$ limit.
		
		In three-dimensions, the spacetime decomposed Poincar\'e algebra takes a simpler form
		\begin{align}
			\left[G_a, H\right] & = P_a \,,  &\left[J, P_a\right] &= -\e_{ab} P^b \,, &\left[J, G_a\right] &= -\e_{ab} G^b \nn\\
			\left[G_a, P_b \right] & = \d_{ab} H \,, & \left[G_a, G_b\right] & = - \e_{ab} J \,,
			\label{3dPoincare}
		\end{align}
		To obtain the extended Bargmann algebra \eqref{3dEBA}, we need to introduce two central generators $M$ and $S$. Then, if we make the following scaling and redefinition
		\begin{align}
			H & \to H + \lambda^2 M\,, & J & \to J + \lambda^2 S \,, & P_a & \to \lambda P_a \,, &  G_a &  \to \lambda G_a \,,
		\end{align}  
		we precisely recover the extended Bargmann algebra. For the next order extended Newtonian algebra \cite{Ozdemir:2019orp}, we need to consider two copies of the Poincar\'e algebra along with two $U(1)$ generators $Y$ and $Z$. In this case, the following definitions precisely recover the extended Newtonian algebra
		\begin{align}
			J & =  J^{1} +  J^{2}\,, & S  &= \l^2 \left( \alpha_1^2 J^{1} +  \alpha_2^2 J^{2}\right) \,,\nn\\
			H &= H^{1} + H^{2} \,, & M & = \lambda^2 \left( \a_1^2 H^{1} + \a_2^2 H^{2} \right) \,,\nn\\
			P_a & = \lambda \left( \a_1 P_a^{1} + \a_2 P_a^{2}  \right) \,, & T_a & = \lambda^3 \left( \a_1^3 P_a^{1} + \a_2^3 P_a^{2}  \right) \,,\nn\\
			G_a & = \lambda \left( \a_1 G_a^{1} + \a_2 G_a^{2}  \right) \,, & B_a & = \lambda^3 \left( \a_1^3 G_a^{1} + \a_2^3 G_a^{2}  \right) \,,\nn\\
			Y &= \lambda^4 \left( \a_1^4 H^{1} + \a_2^4 H^{2} \right) & Z & = \lambda^4 \left( \a_1^4 J^{1} + \a_2^4 J^{2} \right) \,.
		\end{align}
		Once again, these relations can be generalized as
		\begin{align}
			J^{(2n)} & = \lambda^{2n}  \bigoplus\limits_{i=1}^{N_0} \alpha_i^{2n} J^{i} \,, & 	H^{(2n)} & = \lambda^{2n}  \bigoplus\limits_{i=1}^{N_0} \alpha_i^{2n} H^{i} \,, \nn\\
			P_a^{2n+1} &  = \lambda^{2n+1}  \bigoplus\limits_{i=1}^{N_1} \alpha_i^{2n+1} P_{a}^{i} \,, & 	G_a^{2n+1} &  = \lambda^{2n+1}  \bigoplus\limits_{i=1}^{N_1} \alpha_i^{2n+1} G_{a}^{i} \,, 
		\end{align}
		where truncation condition is now $N_0 = N_1 + 1$, and we impose that $\alpha_i \neq 1$ and $\alpha_i \neq \alpha_j$ for $i \neq j$. Note that $J^{2N_1 + 2}$ and $H^{2N_1 +2}$ are the central charges of the extended non-relativistic algebra that we inherit from the $U(1) \times U(1)$ extension of $N$-copies of the Poincar\'e algebra.

	\section{Ultra-Relativistic Scaling Limit of Multimetric Gravity}\label{URChapter}
	\paragraph{}	
	
	The main difference between the non-relativistic and ultra-relativistic expansion is the change of the character of the generator of time translations $H$,  and spatial translations $P_a$. In the ultra-relativistic expansion of the Poincar\'e algebra, $H$ is expanded in odd-powers of $\lambda$ while $P_a$ is expanded in even powers. This is also reflected in the expansion of the corresponding gauge fields, see \eqref{ExpUR}. As a result, the $D$-dimensional Einstein-Hilbert action captures all $\mathcal{O}(\lambda^{2N+1})$ terms, giving rise to an invariant Lagrangian as long as $N_0 = N_1$. This statement is true for $D=3$ as well, hence we do not need to present a separate treatment for $D=3$ but can provide a representative example for $D=4$ which can be generalized to arbitrary dimensions.
	
	In four dimensions, the structure of the spacetime decomposed Einstein-Hilbert action \eqref{4dEH} implies that the ultra-relativistic expansion with \eqref{ExpUR} yields the following structure
	\begin{eqnarray}
		\cL_{\rm EH} &=& \l M^2 \cL_{1} + \l^3 M^2 \cL_{3} + \l^5 M^2 \cL_{5} + \ldots \,.
		\label{URLag}
	\end{eqnarray}
	The lowest order Lagrangian can be isolated by the lowest order expressions for the relativistic fields \eqref{CarrollExp} which can be followed by the rescaling the mass parameter $M^2 \to M^2/\lambda$ and the scaling limit $\lambda \to 0$. To isolate the higher-order Lagrangians, we need to find an $\alpha$-expansion which we discussed previously for the non-relativistic models. Following our footsteps, we start with a non-interacting $N$-metric theory
	\begin{eqnarray}
		\cL &=& \sum_{i=1}^N \left[2M_{(i)}^2 \epsilon_{a b c} \left(R(\Omega^{a}_{(i)})\wedge E^b_{(i)} \wedge E^c_{(i)} - R(\Omega^{a b}_{(i)})\wedge E_{(i)}^c \wedge T_{(i)} \right)  \right] \,,
	\end{eqnarray}
	where $i,j=1,\ldots N$ label the set of fields $\{T_{(i)},E_{(i)}, \Omega^{ab}_{(i)}, \Omega^a_{(i)}\}$.  Then, we keep the expansion of the first set of fields $\{T_{(1)}, E_{(1)}, \Omega^a_{(1)}, \Omega^{ab}_{(1)}\}$ the same as \eqref{ExpUR}
	\begin{align}
		T_{(1)} & = \sum_{n=0}^{N_1} \lambda^{2n+1} \tau_{(2n+1)} \,, & E^a_{(1)} & = \sum_{n=0}^{N_0}  \lambda^{2n} e^a_{(2n)} \,,\nn\\
		\Omega^{ab}_{(1)}  & = \sum_{n=0}^{N_0} \lambda^{2n} 	\omega^{ab}_{(2n)} \,, &  \Omega^a_{(1)} & = \sum_{n=0}^{N_1}  \lambda^{2n+1} \omega^a_{(2n+1)} \,.
		\label{ExpUR1}
	\end{align}
	while for $i > 1$, we incorporate the dimensionless parameters $\alpha_{(i)}$ 
	\begin{align}
		T_{(i)} & = \sum_{n=0}^{N_1} \lambda^{2n+1} \alpha_{(i)}^{2n+1} \tau_{(2n+1)} \,, & E^a_{(i)} & = \sum_{n=0}^{N_0}  \lambda^{2n} \alpha_{(i)}^{2n}  e^a_{(2n)} \,,\nn\\
		\Omega^{ab}_{(i)}  & = \sum_{n=0}^{N_0} \lambda^{2n} \alpha_{(i)}^{2n} 	\omega^{ab}_{(2n)} \,, &  \Omega^a_{(i)} & = \sum_{n=0}^{N_1}  \lambda^{2n+1} \alpha_{(i)}^{2n+1}  \omega^a_{(2n+1)} \,.
		\label{AlphaExpansionUR}
	\end{align}
	where $\alpha_i \neq 1$ and $\alpha_i \neq \alpha_j$ for $i \neq j$. This yield the following expansion of $\cL_{(i)}$ that is the Einstein-Hilbert action for $i$-th set of fields $\{T_{(i)}, E_{(i)}, \Omega^a_{(i)}, \Omega^{ab}_{(i)}\}$ 
	\begin{eqnarray}
		\cL_{EH (i)} &=&  \l \alpha_{i} M_{i}^2 \cL_{1} + \l^3 \alpha_{i}^3 M_{i}^2 \cL_{3} + \l^5 \alpha_{i}^5 M_{i}^2 \cL_{5} + \ldots \,.
	\end{eqnarray}
	Keeping the first set of fields with the standard expansion while using the $\alpha$-expansion for the reminder of the fields, the Einstein-Hilbert action becomes
	\begin{eqnarray}
		\cL_{EH} &=& \sum_{n=0}^{N} \l^{2n+1} \left( M_1^2 + \sum_{i=2}^{N+1} \a_i^{2n+1} M_i^2 \right) \cL_{2n+1} + \mathcal{O}(\l^{2N+3}) \,.
	\end{eqnarray}
	Once again, we isolate the order-$2N+1$ Lagrangian by solving the $N$ number of algebraic equations that determines the values of $\alpha_i$ for $i = 2, 3, \ldots N$, i.e.
	\begin{eqnarray}
		0&=&	 M_1^2 + \sum_{i=2}^{N+1} \a_i^{2n+1} M_i^2 \,, \qquad \text{for} \qquad n = 0,1,\ldots, N-1 \,.
	\end{eqnarray}
	Finally, by rescaling of the mass parameters $M_i^2 \to M_i^2/\lambda^{2N+1}$ and performing the $\lambda \to 0$ limit we eliminate any lower-order divergences and higher order Lagrangians to obtain the desired ultra-relativistic model. Let us now provide examples.
	
	\begin{enumerate}
		\item {\textbf{Carroll Gravity:} The Carroll gravity appears as $\cL_1$ in the ultra-relativistic expansion of Einstein-Hilbert action \eqref{URLag}. It is, thus, sufficient to consider the Poincar\'e algebra, a single set its gauge fields and the Einstein-Hilbert action. Using the lowest order definitions for the relativistic fields 
			\begin{align}
				T &=  \lambda \tau \, & E^a &= e^a \,,&  \Omega^{ab} & = \omega^{ab} \,, & \Omega^a & = \lambda \omega^a \,, & M^2 & = \frac{m^2}{\lambda} \,.
			\end{align}
			we obtain the Carroll gravity 
			\begin{eqnarray}
				\cL_1 &=& -2 \e_{abc} R^{ab}(\o) \wedge e^a \wedge \tau + 2 \e_{abc} R^a(\o) \wedge e^b \wedge e^c \,,
				\label{CarrollGravity}
			\end{eqnarray}
			upon rescaling the mass parameter with $M^2 \to M^2/\l$ and taking the scaling limit $\l \to 0$.
		}
		\item {\textbf{Beyond the Carroll Gravity:} The next order ultra-relativistic model appears at the $\lambda^3$-order in the Lie algebra expansion. Thus, we shall consider two copies of the Einstein-Hilbert action. As for the non-relativistic scaling limit, we keep the same form as the Lie algebra expansion for the first set of fields,
			\begin{align}
				T_{1}  &= \l \tau + \lambda^3 m \,, & E^a_{1}  &=  e^a + \lambda^2 t^a \,,&	\Omega^{ab}_{1}  & = \omega^{ab} + \l^2 s^{ab} \,, & \Omega^a_{1}  & = \lambda \omega^a + \lambda^3 b^a \,.
			\end{align}
			For the second copy, we incorporate the free parameter $\alpha_2$ 
			\begin{align}
				T_{2}  &= \l \tau + \lambda^3 \alpha_2^3 m \,, & E^a_{2}  &=  \alpha_2 e^a + \lambda^2 \alpha_2^2 t^a \,,\nn\\
				\Omega^{ab}_{2}  & = \omega^{ab} + \l^2 \alpha_2^2 s^{ab} \,, & \Omega^a_{2}  & = \lambda \alpha_2 \omega^a + \lambda^3 \alpha_2^3 b^a \,.
			\end{align}
			The combination of these two Lagrangian with mass parameters $M_{1,2}$ gives rise to
			\begin{eqnarray}
				\cL &=& \lambda M^2 \left(1 + \frac12 \alpha_2\right) \cL_1 + \lambda^3 M^2 \left(1+ \frac12 \alpha_2^3\right) \cL_3 + \mathcal{O}(\lambda^5) \,.
			\end{eqnarray}
			where we set $M_1^2 = 2 M_2^2 = M^2$ for simplicity. Here. $\cL_1$ refer to the Carroll gravity \eqref{CarrollGravity} and $\cL_3$  is what we refer to as the beyond the Carroll gravity
			\begin{eqnarray}
				\cL_3 &=&  -2 \e_{abc} R^{ab}(s) \wedge e^c \wedge \tau  -2 \e_{abc} R^{ab}(\o) \wedge t^c \wedge \tau -2 \e_{abc} R^{ab}(\o) \wedge e^c \wedge m \nn\\
				&& + 2 \e_{abc} R^a(b) \wedge e^b \wedge e^c + 4 \e_{abc} R^a(\o) \wedge e^b \wedge t^c  \,.
				\label{BeyondCarrollGravity}
			\end{eqnarray}
			Note that this is the same form of the non-relativistic limit, which is no surprising since in four dimensions, both the non-relativistic and the ultra-relativistic expansion of the Einstein-Hilbert action take the same form in series expansion in $\lambda$, see \eqref{NREH} and \eqref{URLag}. We remind the reader that this is not longer true for $D \neq 4$. 
			Nevertheless, we can fix $\alpha_2$ to be $\alpha_2 = - 2$ which cancels out the coefficient of the Carroll gravity and fixed the coefficient of $\cL_3$ to be $-3$.  Then, the following scaling of the mass parameter
			\begin{eqnarray}
				M^2 = \frac{1}{3 \lambda^3} m^2 \,.
			\end{eqnarray}
			along with the scaling limit $\lambda \to 0$ recover the beyond the Carroll gravity model up to an overall minus sign.
		}
	\end{enumerate}

The multi-gravity models that we discussed so far needs to include a potential term for the vielbeine to be physically viable. Unlike the non-relativistic case, however, these terms appear at the $\mathcal{O}(\l)$ Lagrangian due to the changing expansion character of the spatial and the temporal vielbein
\begin{eqnarray}
	\cL_{\rm{pot}} &=& \l \left( \e_{abc}  \tau \wedge e^a \wedge e^b \wedge e^c \right) \nn\\
	&& + \lambda^3 \left( e_{abc}  m \wedge e^a \wedge e^b \wedge e^c + 3 e_{abc}  \tau \wedge e^a \wedge e^b \wedge t^c \right)  + \mathcal{O}(\l^5)\,.
\end{eqnarray}
This is precisely the same expansion of the non-relativistic models except that the Lagrangians are now shifted by $\lambda^{-2}$. Nevertheless, our arguments for the elimination of the lower-order cosmological terms still hold since the structure of the potential terms, \eqref{CosmologicalConstant}, contains more than necessary the number of free coefficients.

	We end this section with a brief discussion on how to obtain the ultra-relativistic higher-order algebras by contracting the multiple copies of the Poincar\'e algebra. As with the non-relativistic algebras for $D \geq 4$, the ultra-relativistic algebras that admit an invariant action formulation arise from the Lie algebra expansion for $N_0  = N_1$. This implies that for an algebra of order $(N_1,N_1)$, one has the same number of generators as $(N_1 + 1)$-copies of Poincar\'e algebra. Furthermore, the expansion does not change the structure constant of the smaller core algebra, indicating that we can combine the generators of multiple copies of the Poincar\'e algebra in a linearly independent way to exhibit the generators of the larger non/ultra-relativistic algebras. Based on the $\alpha$-expansion of the relativistic fields, we introduce the following expressions for the ultra-relativistic generators in terms of the generators of the Poincar\'e algebra
	\begin{align}
		J_{ab}^{(2n)} & = \lambda^{2n}  \bigoplus\limits_{i=1}^{N_0} \alpha_i^{2n} J_{ab}^{i} \,, & 	H^{(2n+1)} & = \lambda^{2n+1}  \bigoplus\limits_{i=1}^{N_1} \alpha_i^{2n+1} H^{i} \,, \nn\\
		P_a^{(2n)} &  = \lambda^{2n}  \bigoplus\limits_{i=1}^{N_0} \alpha_i^{2n} P_{a}^{i} \,, & 	G_a^{(2n+1)} &  = \lambda^{2n+1}  \bigoplus\limits_{i=1}^{N_1} \alpha_i^{2n+1} G_{a}^{i} \,, 
	\end{align}
	where $\alpha_i \neq 1$ and $\alpha_i \neq \alpha_j$ for $i \neq j$. Note that in this case these alpha parameters are free and not necessary to be fixed.  This direct sum structure trivially satisfy the algebra \eqref{URInfinite}. Once again, these relations can be inverted to express the relativistic generators in terms of the ultra-relativistic ones. Furthermore, we remind the reader that the truncation condition is $N_0 = N_1$.

		\section{Discussion}\label{Discussion}
			\paragraph{}
		In this work, we have presented non-relativistic and ultra-relativistic scaling limits of multimetric gravity theories. By the field content of the multimetric gravity, it is expected that these models contain more number of degrees of freedom compared to that of Galilei/Carroll gravity, which are the limits of General Relativity. We have shown that the limits of multimetric gravity correspond to non/ultra-relativistic gravity with extended symmetries. In particular, we have shown that the non-relativistic limit of bi-metric gravity is the recent formulation of an action principle for the Newtonian gravity when no potential terms are present. On the other hand, turning on the potential terms yield a constant background mass density in the non-relativistic sector. We expect that the scaling limit that we provide in this paper will be helpful in phenomenological studies for both the multigravity and the non/ultra-relativistic gravity. Especially, given the fact that the Newtonian gravity has a spherically symmetric vacuum solution \cite{VandenBleeken:2019gqa, Ergen:2020yop, Hansen:2019vqf, Hansen:2020pqs} and passes the three classical tests of General Relativity, it would be interesting to see the bimetric origin of this solution and relevant phenomenological consequences.
		
		The work we present can be regarded as a starting point for various further studies. On the technical side, it would be interesting to extend the analysis that we presented here for the string limit and corresponding algebras, i.e., string limit of Einstein-Hilbert action,  see \cite{Bergshoeff:2019ctr,Bergshoeff:2018vfn}. Furthermore, here we focus on three and higher dimensions. The reason for this is that the two-dimensional gravitational theories in their $BF$-formalism cannot be established just by the gauge fields of the Poincar\'e (or (A)dS) algebra, but they require the presence of matter fields transforming in the coadjoint representation. We believe that the same scaling limits can be defined in the presence of the matter sector, however, a careful analysis is indeed essential.
		
		A rather interesting continuation of our work would be to include supersymmetry. Although the non/ultra-relativistic superalgebras are now well-understood, thanks to the Lie algebra expansion, we also need theories that contain both gauge fields and matter fields. At present, the existing techniques, which are based on the Lie algebra expansion, can produce reducible representations for the matter multiplets of supersymmetric theories \cite{Kasikci:2021atn}. Furthermore, these multiplets are rigid and their extensions to local theories have been an open problem. We hope that the scaling limit that we present in this paper can shed light on this open problem.
		
		Another interesting point is that although we have achieved the scaling limit with multiple copies of the Poincar\'e algebra, the same limit can also be established with the Poincar\'e and the Euclidean algebras, see \cite{Ozdemir:2019orp} for a three-dimensional example. When the Euclidean algebra is considered, the difference arises in the commutation relations that include $H \equiv P_0$ and $G_a \equiv J_{0a}$, i.e.
		\begin{align}
			\left[G_a, P_b\right] &= \d_{a b} H\,, &  \left[G_a, H\right] & = - P_a \,, &\left[J_{a b}, P_c\right] &= \d_{b c} P_a - \d_{a c} P_b\,, \nn\\
			\left[J_{a b}, G_{c}\right] &= \d_{b c} G_a - \d_{a c} G_b \,, &\left[J_{a b}, J_{c d}\right] &= 4 \d_{[a[c}J_{d]b]}\,, & \left[G_a, G_b\right] & = - J_{a b}\,. 
			\label{EuclideanAlgebra}
		\end{align}
	In the contraction process to obtain the non/ultra-relativistic algebras, the minus factor that appears in the $[G_a, H]$ and $[G_a, G_b]$ commutators can be handled by introducing a minus sign in the definitions of the non/ultra-relativistic generators of extended algebras. For instance, in the case of non-relativistic extended algebras, we have the following definitions for the generators 
	\begin{align}
J^{(2n)}_{a b} & = \lambda^{2n} \bigoplus\limits_{i=1}^{N_0}  \alpha_i^{2n}  \sigma_i^{n} J^i_{a b} \,, &   H^{(2n)} & = \lambda^{2n} \bigoplus\limits_{i=1}^{N_0}  \alpha_i^{2n}  \sigma_i^{n} H^i   \nn \\
P^{(2n+1)}_{a} & = \lambda^{2n+1} \bigoplus\limits_{i=1}^{N_1}  \alpha_i^{2n+1}  \sigma_i^{n+1} P^i_{a} \,, & G^{(2n+1)}_{a} & = \lambda^{2n+1} \bigoplus\limits_{i=1}^{N_1}  \alpha_i^{2n+1}  \sigma_i^{n} G^i_{a}\,,
\end{align}
	where $\alpha_i \neq 1$ and $\alpha_i \neq \alpha_j$ for $i \neq j$. The necessary sign is introduced by the $\sigma_i$ which is defined as $\sigma_i=1$ for Lorentizan and $\sigma_i=-1$ for Euclidean algebras. With this relation in hand, it is tempting to propose a bimetric theory of gravity that is the sum of Einstein-Hilbert action in the Lorentizan and the Euclidean signatures as a simple, worthwhile phenomenological model. To our knowledge, such a model has not been investigated in the literature. We hope to study the phenomenological aspects of this model in a near future. The multiple copy structure of the Poincar\'e algebra also suggest a possible connection with our construction and colored gravity \cite{Gwak:2015vfb,Gwak:2015jdo}, which may be another interesting avenue to investigate.
	
	As an important remark, in Appendix \ref{AppA}, we have explicitly shown the equivalence of the first and the second order formulation of the Newtonian gravity. It would be interesting to see if the same is true between the ultra-relativistic Beyond Carroll gravity action \eqref{BeyondCarrollGravity} and the metric formulation \cite{Hansen:2021fxi}. Finally, the coadjoint Poincar\'e (or AdS) algebra, which is a Lie algebra expansion of the Poincar\'e (or AdS) algebra for $\{P_A, J_{AB}\} \subset V_0$, is also known to reproduce certain three and four-dimensional extended non-relativistic algebras \cite{Bergshoeff:2020fiz}. It would be interesting to see if there is a relation between the coadjoint algebras and multiple copies of Poincar\'e and Euclidean algebra.
	
	\vspace{0.5cm}
	 {\bf Acknowledgements}
		We thank Eric Bergshoeff and Johannes Lahnsteiner for discussions. The work of O.K. and C.B.S. is supported by TUBITAK grant 121F064. M.O. is supported in part by TUBITAK grant 121F064 and Istanbul Technical University Research Fund under grant number TGA-2020-42570. M.O. acknowledges the support by the Distinguished Young Scientist Award BAGEP of the Science Academy. M.O. also acknowledges the support by the Outstanding Young Scientist Award of the Turkish Academy of Sciences (TUBA-GEBIP). U.Z. is supported by TUBITAK - 2218 National Postdoctoral Research Fellowship Program with grant number 118C512.

		\appendix
		
				\section{Equivalence of the First Order and the Second Order Formulation of the Newtonian Gravity}\label{AppA}
				\paragraph{}
	In this Appendix, we show that the first order formulation of Newtonian gravity that arises from the Lie algebra expansion is equivalent to the second order formulation that arises from the $1/c^2$ expansion of the General Relativity. The first order formulation of Newtonian gravity appears at $\mathcal{O}(\lambda^3)$ in the non-relativistic expansion of the Einstein-Hilbert action and is given as
		\begin{eqnarray}
				S_{N} &=& \int d^4x\ \epsilon^{\mu \nu \rho \sigma} \epsilon_{abc} \Big( e_\mu{}^a e_\nu{}^b R_{\r\s}{}^c(\o) -   e_\mu{}^a \t_\nu R_{\r\s}{}^{bc}(s)  - e_\mu{}^a m_\nu R_{\r\s}{}^{bc}(\o)  \nn\\
				&& \qquad \qquad  \qquad\quad - t_\mu{}^a \t_\nu R_{\r\s}{}^{bc}(\omega) \Big)\,, \label{eric3order}
		\end{eqnarray}
	where the group-theoretical curvatures are given in \eqref{NewtonianCurvatures}. Using the inverse spatial and the temporal vielbein, this action can be put in a more useful form
	\begin{eqnarray}
		S_N &=& \int d^4x\ e \ \bigg( - 4 \tau^\mu e^\nu{}_a R_{\m\n}{}^a (\o) +    2 e^\mu{}_a  e^\nu{}_b R_{\m\n}{}^{ab}(s)  - 2 \t^\r m_\r  e^\mu{}_a  e^\nu{}_b R_{\m\n}{}^{ab}(\omega)\, \nn \\
		&&   + 4  e^\r{}_a m_\r  \t^\mu  e^\nu{}_b R_{\m\n}{}^{ab}(\omega) +  2  e^\r{}_c t_\r{}^c  e^\mu{}_a  e^\nu{}_b R_{\m\n}{}^{ab}(\omega)  - 4 e^\r{}_a t_\r{}^b  e^\mu{}_b  e^\nu{}_c R_{\m\n}{}^{ac}(\omega) \bigg)\,, \quad \label{eric3order3}
	\end{eqnarray} 
where $e = \det (\tau_\mu, 	e_\mu{}^a)$. Note that we made a slight change of notation $\omega^a \rightarrow  - \omega^a\,,   s^{ab} \rightarrow  - s^{ab}$	to match with the conventions of \cite{Hansen:2019pkl}. The inverse elements satisfy the following properties
\begin{align}
	\tau^\mu e_\mu{}^a & = 0 \,, & \tau_\mu e^\mu{}_a & = 0 \,, & \tau^\mu \tau_\mu& = -1 \,, \nn\\
	e_\mu{}^a e^\mu{}_b & = \delta^a{}_b \,, & e_\mu{}^a e^\nu{}_a & = \delta_\mu{}^\nu + \tau_\mu \tau^\nu \,.
\end{align}
Note here that the inverse temporal vielbein have non-trivial Galilean transformations. However, a Galilean invariant timelike vector can be constructed by using the gauge field $m_\mu$ and vielbein as $\hat{\t}^\mu = \t^\mu-  h^{\m\n}m_\n$. Note that $\hat \tau^\mu$ also satisfies the invertibility conditions. To show the equivalence of the first and the second order formulation, we first to use the following Bianchi identity 
		\bea
		\tau\wedge R^a(\omega)- e_b \wedge R^{ab}(\omega) = 0\,, \label{bianchiId}
		\eea 
		which is a consequence of the curvature constraints that are necessary to express $\omega_\mu{}^{ab}, \omega_\mu{}^a$ in terms of the independent fields of the Newtonian gravity $\{e_\mu{}^a\,, \tau_\mu\,, m_\mu\}$  \cite{Andringa:2010it}. Utilizing the Bianchi identity and using the Galilean invariant timelike vector, the first-order action becomes
			\bea
		S_{N} &=& - \frac{1}{32\pi G} \int d^4x\ e \ \bigg( - 4 \hat{\t}^\mu e^\nu{}_aR_{\m\n}{}^a(\omega)+    2 e^\mu{}_a  e^\nu{}_b R_{\m\n}{}^{ab}(s)  - 2 \t^\r m_\r  e^\mu{}_a  e^\nu{}_b R_{\m\n}{}^{ab}(\omega)\, \nn \\
		&& \qquad \qquad \qquad  \qquad +  2  e^\r{}_c t_\r{}^c  e^\mu{}_a  e^\nu{}_b R_{\m\n}{}^{ab}(\omega)  - 4 e^\r{}_a t_\r{}^b  e^\mu{}_b  e^\nu{}_c R_{\m\n}{}^{ac}(\omega) \bigg)\,. \label{eric3order4}
		\eea 
		where we introduce proper numerical prefactor and the Newton's constant for future purposes. The Riemann tensor is then given by \cite{Andringa:2010it}
		\bea
		R_{\m\n\s}{}^{\r} =  e^\rho{}_a\t_\s R_{\m\nu}{}^a(\omega) - e_{\s a}e^\r{}_b R_{\mu\nu}{}^{ab}(\omega)\,, \label{Riemann}
		\eea 
		from which we read-off the following contraction that is an important quantity when relating the first and the second order formulations
		\bea
		- \hat{\t}^\mu e^\nu{}_aR_{\m\n}{}^a(\omega) &=& \hat{\t}^\mu  \hat{\t}^\nu  \bar{R}_{\m\n}\,. \label{Romega}
		\eea 
		Here, we use the $ \bar{R}_{\m\n}$ notation in order to indicate that the connection is the Galilean invariant to stay compatible with the notation of \cite{Hansen:2019pkl}. With these results in hand, let us focus back to the action \eqref{eric3order4}, in particular the term that involve the curvature of $s^{ab}$
		\bea
		\epsilon^{\mu \nu \rho \sigma} \epsilon_{abc}  e_\mu{}^a \t_\nu R_{\r\s}{}^{bc}(s) &=& \epsilon^{\mu \nu \rho \sigma} \epsilon_{abc}  e_\mu{}^a \t_\nu (2\partial_\r s_\s{}^{bc}  + 2  \omega_{\r}{}^{bd}s_{\s d}{}^{c} - 2  \omega_{\r}{}^{cd}s_{\s d}{}^{b} \nn\\
		&&   - \omega_{\r}{}^b \omega_{\s}{}^c  +  \omega_{\s}{}^b \omega_{\r}{}^c)\,. 
		\eea
		The integration by parts of the derivative term yield the curvature constraints for $\tau_\mu$ and $e_\mu{}^a$ which are already imposed. Thus, what remains is the last two terms that involve $\omega^a$  
		\bea
		\epsilon^{\mu \nu \rho \sigma} \epsilon_{abc}  e_\mu{}^a \t_\nu \left( - \omega_{\r}{}^b \omega_{\s}{}^c  +  \omega_{\s}{}^b \omega_{\r}{}^c \right) =  - 2 e \ \left( \nabla_{\mu} \t^\mu \nabla_{\nu} \t^\nu  -  \nabla_{\mu} \t^\nu \nabla_{\nu} \t^\mu  \right)\,.  \label{omegaomega}
		\eea 
		which can be expressed as follows after partial integration 
		\bea
		- 2 e \ \left( \nabla_{\mu} \t^\mu \nabla_{\nu} \t^\nu  -  \nabla_{\mu} \t^\nu \nabla_{\nu} \t^\mu  \right) &=&-2 \hat{\t}^\mu \hat{\t}^\nu  \bar{R}_{\mu\nu}  - 2 h^{\mu\sigma} h^{\l\r} m_\s m_\l R_{\m\r}\,, \label{omegaterm}
		\eea 
		Here, we used the definition $\hat{\t}^\mu$ as well as the identities \cite{Hansen:2020pqs}
		\begin{align}
		\left[\nabla_\mu, \nabla_\nu \right]  X_\sigma &= - R_{\mu\nu\sigma}{}^\rho X_\rho \,, &	\bar R_{\mu\nu\lambda}{}^\rho h^{\lambda\sigma} &= 	\bar R_{\mu\nu\lambda}{}^\sigma h^{\lambda\rho}  \,,
		\end{align}
		where $h^{\mu\nu} = e^{\mu}{}_a e^{\nu \, a}$. Note that we assume the vanishing of torsion since it is not required to establish the action principle that gives rise to the Poisson equation for Newtonian gravity as an equation of motion. Finally, focusing on the $R^{ab}(\omega)$ terms in the Lagrangian (\ref{eric3order4}), we have
		\begin{align}
  - 2 \Phi h^{\mu\nu}\bar{R}_{\mu\nu}  -  \Phi_{\m\nu}  h^{\mu\nu}  h^{\r\s}\bar{R}_{\r\s} + 2 \Phi_{\mu\l}h^{\mu\s} h^{\nu\l} \bar{R}_{\n\s} & =  2 \t^\r m_\r  e^\mu{}_a  e^\nu{}_b R_{\m\n}{}^{ab}(\omega) \nn\\
  & +  2  e^\r{}_c t_\r{}^c  e^\mu{}_a  e^\nu{}_b R_{\m\n}{}^{ab}(\omega) \nn\\
& - 4 e^\r{}_a t_\r{}^b  e^\mu{}_b  e^\nu{}_c R_{\m\n}{}^{ac}(\omega)  \,, 
		\end{align}
		where we used the following definitions
		\bea
		\Phi = - \t^\m m_{\m}\,, \qquad 	\Phi_{\m\n} = \delta_{ab} \left( e_\mu{}^a t_\nu{}^b  + e_\nu{}^a t_\mu{}^b   \right)\,.
		\eea 
		We now introduce the Galilean invariant definitions of $\Phi$ and $\Phi_{\m\nu}$ as adding and subtracting the appropriate terms as
	\begin{align}
	- 2 \Phi h^{\mu\nu}\bar{R}_{\mu\nu} - \Phi_{\m\nu}  h^{\mu\nu}  h^{\r\s}\bar{R}_{\r\s}  + 2 \Phi_{\mu\l}h^{\mu\s} h^{\nu\l} \bar{R}_{\n\s} &=  2 \bar{\Phi}_{\mu\l}h^{\mu\s} h^{\nu\l} \bar{R}_{\n\s} - \bar{\Phi}_{\m\nu}  h^{\mu\nu}  h^{\r\s}\bar{R}_{\r\s}    \nn\\
& - 2 \bar{\Phi}  h^{\mu\nu}\bar{R}_{\mu\nu}  + 2   h^{\mu\sigma} h^{\nu\r} m_\s m_\r R_{\m\n}\,, \label{phibar}
	\end{align}
		where the Galilean invariant quantities are defined as \cite{Hansen:2020pqs},
		\bea
		\bar{\Phi} =  \Phi  + \frac{1}{2} h^{\m\n} m_\m m_\n\,, \qquad 	\bar{\Phi}_{\m\n} =  \Phi _{\m\n} -  m_\m m_\n\,. 
		\eea 
		Substituting these results back in the action (\ref{eric3order4}), we obtain 
		\begin{eqnarray}
			S_N
			&=&   \frac{1}{16\pi G} \int d^4x\ e \ \bigg( - \hat{\t}^\mu \hat{\t}^\nu  \bar{R}_{\mu\nu}  +  \bar{\Phi}  h^{\mu\nu}\bar{R}_{\mu\nu}  - \bar{\Phi}_{\r\s}h^{\mu\r} h^{\nu\s} (\bar{R}_{\m\n} - \frac{1}{2} h_{\m\n} h^{\k\l}   \bar{R}_{\k\l} ) \bigg)\,, \qquad \label{Newtonianaction}  \,,
		\end{eqnarray}
		which is the  action principle that gives rise to the Poisson equation for Newtonian gravity as an equation of motion \cite{Hansen:2019pkl}. 
		
		As mentioned in Section \ref{NR4}, we can add a cosmological constant to this action $e \Lambda$, which arise from the contraction of the bimetric gravity with a non-vanishing potential terms for vielbein. This would give rise to the Poisson equation as an equation of motion \cite{Hansen:2019pkl}
		\begin{eqnarray}
		\nabla^2 \Phi &=& 4 \pi G \Lambda\,,
		\end{eqnarray}
	 which represents a constant background mass density.


\begin{thebibliography}{99}
			
			\bibitem{Hatsuda:2001pp}
			M.~Hatsuda and M.~Sakaguchi,
			``Wess-Zumino term for the AdS superstring and generalized Inonu-Wigner contraction,''
			Prog. Theor. Phys. \textbf{109}, 853-867 (2003)
			
			\bibitem{deAzcarraga:2002xi}
			J.~A.~de Azcarraga, J.~M.~Izquierdo, M.~Picon and O.~Varela,
			``Generating Lie and gauge free differential (super)algebras by expanding Maurer-Cartan forms and Chern-Simons supergravity,''
			Nucl. Phys. B \textbf{662}, 185-219 (2003)
			
			\bibitem{deAzcarraga:2007et}
			J.~A.~de Azcarraga, J.~M.~Izquierdo, M.~Picon and O.~Varela,
			``Expansions of algebras and superalgebras and some applications,''
			Int. J. Theor. Phys. \textbf{46}, 2738-2752 (2007)
			
			\bibitem{Bergshoeff:2019ctr}
			E.~Bergshoeff, J.~M.~Izquierdo, T.~Ort\'\i{}n and L.~Romano,
			``Lie Algebra Expansions and Actions for Non-Relativistic Gravity,''
			JHEP \textbf{08}, 048 (2019)
			
			\bibitem{Bergshoeff:2021tbz}
			E.~A.~Bergshoeff, M.~Ozkan and M.~S.~Zog,
			``The holographic c-theorem and infinite-dimensional Lie algebras,''
			JHEP \textbf{01}, 010 (2022)
			
			\bibitem{VandenBleeken:2017rij}
			D.~Van den Bleeken,
			``Torsional Newton\textendash{}Cartan gravity from the large c expansion of general relativity,''
			Class. Quant. Grav. \textbf{34}, no.18, 185004 (2017)
			
			\bibitem{Hansen:2019pkl}
			D.~Hansen, J.~Hartong and N.~A.~Obers,
			``Action Principle for Newtonian Gravity,''
			Phys. Rev. Lett. \textbf{122}, no.6, 061106 (2019)
			
			  \bibitem{Papageorgiou:2009zc}
    G.~Papageorgiou and B.~J.~Schroers,
    ``A Chern-Simons approach to Galilean quantum gravity in 2+1 dimensions,''
    JHEP \textbf{11} (2009), 009
    doi:10.1088/1126-6708/2009/11/009
     [arXiv:0907.2880 [hep-th]].

\bibitem{Bergshoeff:2016lwr}
E.~A.~Bergshoeff and J.~Rosseel,
``Three-Dimensional Extended Bargmann Supergravity,''
Phys. Rev. Lett. \textbf{116} (2016) no.25, 251601
doi:10.1103/PhysRevLett.116.251601
[arXiv:1604.08042 [hep-th]].

\bibitem{Hartong:2016yrf}
J.~Hartong, Y.~Lei and N.~A.~Obers,
``Nonrelativistic Chern-Simons theories and three-dimensional Ho\v{r}ava-Lifshitz gravity,''
Phys. Rev. D \textbf{94} (2016) no.6, 065027
doi:10.1103/PhysRevD.94.065027
[arXiv:1604.08054 [hep-th]].

\bibitem{Aviles:2018jzw}
L.~Avil\'es, E.~Frodden, J.~Gomis, D.~Hidalgo and J.~Zanelli,
``Non-Relativistic Maxwell Chern-Simons Gravity,''
JHEP \textbf{05} (2018), 047
doi:10.1007/JHEP05(2018)047
[arXiv:1802.08453 [hep-th]].
			
			
			
			
			\bibitem{Ozdemir:2019orp}
			N.~Ozdemir, M.~Ozkan, O.~Tunca and U.~Zorba,
			``Three-Dimensional Extended Newtonian (Super)Gravity,''
			JHEP \textbf{05}, 130 (2019)
			
			
			\bibitem{deAzcarraga:2019mdn}
			J.~A.~de Azc\'arraga, D.~G\'utiez and J.~M.~Izquierdo,
			``Extended $D = 3$ Bargmann supergravity from a Lie algebra expansion,''
			Nucl. Phys. B \textbf{946}, 114706 (2019)
			
			
			\bibitem{Concha:2019lhn}
			P.~Concha and E.~Rodr\'\i{}guez,
			``Non-Relativistic Gravity Theory based on an Enlargement of the Extended Bargmann Algebra,''
			JHEP \textbf{07}, 085 (2019)
			
			\bibitem{Penafiel:2019czp}
			D.~M.~Pe\~nafiel and P.~Salgado-Rebolledo,
			``Non-relativistic symmetries in three space-time dimensions and the Nappi-Witten algebra,''
			Phys. Lett. B \textbf{798}, 135005 (2019)
			
			\bibitem{Gomis:2019fdh}
			J.~Gomis, A.~Kleinschmidt and J.~Palmkvist,
			``Galilean free Lie algebras,''
			JHEP \textbf{09}, 109 (2019)
			
			\bibitem{Ozdemir:2019tby}
			N.~Ozdemir, M.~Ozkan and U.~Zorba,
			``Three-dimensional extended Lifshitz, Schr\"odinger and Newton-Hooke supergravity,''
			JHEP \textbf{11}, 052 (2019)
			
			\bibitem{Gomis:2019sqv}
			J.~Gomis, A.~Kleinschmidt, J.~Palmkvist and P.~Salgado-Rebolledo,
			``Symmetries of post-Galilean expansions,''
			Phys. Rev. Lett. \textbf{124}, no.8, 081602 (2020)
			
			\bibitem{Gomis:2019nih}
			J.~Gomis, A.~Kleinschmidt, J.~Palmkvist and P.~Salgado-Rebolledo,
			``Newton-Hooke/Carrollian expansions of (A)dS and Chern-Simons gravity,''
			JHEP \textbf{02}, 009 (2020)
			
			\bibitem{Kasikci:2020qsj}
			O.~Kasikci, N.~Ozdemir, M.~Ozkan and U.~Zorba,
			``Three-dimensional higher-order Schr\"odinger algebras and Lie algebra expansions,''
			JHEP \textbf{04}, 067 (2020)
			
			\bibitem{Concha:2020sjt}
			P.~Concha, M.~Ipinza and E.~Rodr\'\i{}guez,
			``Generalized Maxwellian exotic Bargmann gravity theory in three spacetime dimensions,''
			Phys. Lett. B \textbf{807}, 135593 (2020)
			
			\bibitem{Concha:2020ebl}
			P.~Concha, L.~Ravera, E.~Rodr\'\i{}guez and G.~Rubio,
			``Three-dimensional Maxwellian Extended Newtonian gravity and flat limit,''
			JHEP \textbf{10}, 181 (2020)
			
			\bibitem{Concha:2020eam}
			P.~Concha, M.~Ipinza, L.~Ravera and E.~Rodr\'\i{}guez,
			``Non-relativistic three-dimensional supergravity theories and semigroup expansion method,''
			JHEP \textbf{02}, 094 (2021)
			
			

\bibitem{Concha:2020tqx}
P.~Concha, L.~Ravera and E.~Rodr\'\i{}guez,
``Three-dimensional non-relativistic extended supergravity with cosmological constant,''
Eur. Phys. J. C \textbf{80} (2020) no.12, 1105
doi:10.1140/epjc/s10052-020-08685-2
[arXiv:2008.08655 [hep-th]].

			
			\bibitem{Concha:2021jos}
			P.~Concha, L.~Ravera and E.~Rodr\'\i{}guez,
			``Three-dimensional exotic Newtonian supergravity theory with cosmological constant,''
			Eur. Phys. J. C \textbf{81}, no.7, 646 (2021)
			
			\bibitem{Gomis:2022spp}
			J.~Gomis and A.~Kleinschmidt,
			``Infinite-dimensional algebras as extensions of kinematic algebras,''
			[arXiv:2202.05026 [hep-th]].
			
			\bibitem{Grumiller2020}
            D.~Grumiller, J.~Hartong, S.~Prohazka and J.~Salzer,
            ``Limits of JT gravity,''
            JHEP \textbf{02} (2021), 134
            doi:10.1007/JHEP02(2021)134
            [arXiv:2011.13870 [hep-th]].
			
            \bibitem{Gomis2020}
            J.~Gomis, D.~Hidalgo and P.~Salgado-Rebolledo,
            ``Non-relativistic and Carrollian limits of Jackiw-Teitelboim gravity,''
            JHEP \textbf{05} (2021), 162
            doi:10.1007/JHEP05(2021)162
            [arXiv:2011.15053 [hep-th]].

			\bibitem{Ravera:2022buz}
			L.~Ravera and U.~Zorba,
			``Carrollian and Non-relativistic Jackiw-Teitelboim Supergravity,''
			[arXiv:2204.09643 [hep-th]].
			
			\bibitem{Concha:2022you}
			P.~Concha, E.~Rodr\'\i{}guez, G.~Rubio and P.~Ya\~nez,
			``Three-dimensional Newtonian gravity with cosmological constant and torsion,''
			[arXiv:2204.11763 [hep-th]].
			
			
\bibitem{Concha:2021llq}
P.~Concha, L.~Ravera and E.~Rodr\'\i{}guez,
``Three-dimensional non-relativistic supergravity and torsion,''
Eur. Phys. J. C \textbf{82} (2022) no.3, 220
doi:10.1140/epjc/s10052-022-10183-6
[arXiv:2112.05902 [hep-th]].


			\bibitem{Ravera:2019ize}
			L.~Ravera,
``AdS Carroll Chern-Simons supergravity in 2 + 1 dimensions and its flat limit,''
Phys. Lett. B \textbf{795} (2019), 331-338
doi:10.1016/j.physletb.2019.06.026
[arXiv:1905.00766 [hep-th]].


\bibitem{Ali:2019jjp}
F.~Ali and L.~Ravera,
``$\mathcal{N}$-extended Chern-Simons Carrollian supergravities in $2+1$ spacetime dimensions,''
JHEP \textbf{02} (2020), 128
doi:10.1007/JHEP02(2020)128
[arXiv:1912.04172 [hep-th]].


\bibitem{Concha:2021jnn}
P.~Concha, D.~Pe\~nafiel, L.~Ravera and E.~Rodr\'\i{}guez,
``Three-dimensional Maxwellian Carroll gravity theory and the cosmological constant,''
Phys. Lett. B \textbf{823} (2021), 136735
doi:10.1016/j.physletb.2021.136735
[arXiv:2107.05716 [hep-th]].

\bibitem{BergshoeffCarroll}
E.~Bergshoeff, D.~Grumiller, S.~Prohazka and J.~Rosseel,
``Three-dimensional Spin-3 Theories Based on General Kinematical Algebras,''
JHEP \textbf{01} (2017), 114
doi:10.1007/JHEP01(2017)114
[arXiv:1612.02277 [hep-th]].

\bibitem{MatulichCarroll}
J.~Matulich, S.~Prohazka and J.~Salzer,
``Limits of three-dimensional gravity and metric kinematical Lie algebras in any dimension,''
JHEP \textbf{07} (2019), 118
doi:10.1007/JHEP07(2019)118
[arXiv:1903.09165 [hep-th]].

			
			\bibitem{Paulos:2012xe}
			M.~F.~Paulos and A.~J.~Tolley,
			``Massive Gravity Theories and limits of Ghost-free Bigravity models,''
			JHEP \textbf{09}, 002 (2012)
			
\bibitem{Afshar:2014dta}
H.~R.~Afshar, E.~A.~Bergshoeff and W.~Merbis,
``Interacting spin-2 fields in three dimensions,''
JHEP \textbf{01}, 040 (2015)

			
			\bibitem{Bergshoeff:2013xma}
			E.~A.~Bergshoeff, S.~de Haan, O.~Hohm, W.~Merbis and P.~K.~Townsend,
			``Zwei-Dreibein Gravity: A Two-Frame-Field Model of 3D Massive Gravity,''
			Phys. Rev. Lett. \textbf{111}, no.11, 111102 (2013)
			[erratum: Phys. Rev. Lett. \textbf{111}, no.25, 259902 (2013)]
			
			\bibitem{Ozkan:2019iga}
			M.~Ozkan, Y.~Pang and U.~Zorba,
			``Unitary Extension of Exotic Massive 3D Gravity from Bigravity,''
			Phys. Rev. Lett. \textbf{123}, no.3, 031303 (2019)
			doi:10.1103/PhysRevLett.123.031303
			
\bibitem{Sevim:2019scg}
S.~Sevim and M.~S.~Z\"o\u{g},
``Unitarity flow in 2+1 dimensional massive gravity,''
Phys. Rev. D \textbf{102}, no.6, 064050 (2020)


      \bibitem{VandenBleeken:2019gqa}
       D.~Van den Bleeken,
      ``Torsional Newton-Cartan gravity and strong gravitational fields,''
       [arXiv:1903.10682 [gr-qc]].
			\bibitem{Hansen:2019vqf}
			D.~Hansen, J.~Hartong and N.~A.~Obers,
			``Gravity between Newton and Einstein,''
			Int. J. Mod. Phys. D \textbf{28}, no.14, 1944010 (2019)
			
\bibitem{Ergen:2020yop}
M.~Ergen, E.~Hamamci and D.~Van den Bleeken,
Eur. Phys. J. C \textbf{80} (2020) no.6, 563
[erratum: Eur. Phys. J. C \textbf{80} (2020) no.7, 657]
doi:10.1140/epjc/s10052-020-8112-6
[arXiv:2002.02688 [gr-qc]].
		
		 	
			
			\bibitem{Hansen:2020pqs}
			D.~Hansen, J.~Hartong and N.~A.~Obers,
			``Non-Relativistic Gravity and its Coupling to Matter,''
			JHEP \textbf{06}, 145 (2020)
			
			
		
			
			\bibitem{Kasikci:2021atn}
			O.~Kasikci and M.~Ozkan,
			``Lie algebra expansions, non-relativistic matter multiplets and actions,''
			JHEP \textbf{01}, 081 (2022)
			
			
			\bibitem{Bergshoeff:2009hq}
			E.~A.~Bergshoeff, O.~Hohm and P.~K.~Townsend,
			``Massive Gravity in Three Dimensions,''
			Phys. Rev. Lett. \textbf{102}, 201301 (2009)
			
			\bibitem{Sinha:2010ai}
			A.~Sinha,
			``On the new massive gravity and AdS/CFT,''
			JHEP \textbf{06}, 061 (2010)
			
			\bibitem{Paulos:2010ke}
			M.~F.~Paulos,
			``New massive gravity extended with an arbitrary number of curvature corrections,''
			Phys. Rev. D \textbf{82}, 084042 (2010)
			
			
			
	
			
	


\bibitem{Bergshoeff:2014jla}
E.~Bergshoeff, J.~Gomis and G.~Longhi,
``Dynamics of Carroll Particles,''
Class. Quant. Grav. \textbf{31} (2014) no.20, 205009
doi:10.1088/0264-9381/31/20/205009
[arXiv:1405.2264 [hep-th]].

\bibitem{Duval:2014uoa}
C.~Duval, G.~W.~Gibbons, P.~A.~Horvathy and P.~M.~Zhang,
``Carroll versus Newton and Galilei: two dual non-Einsteinian concepts of time,''
Class. Quant. Grav. \textbf{31} (2014), 085016
doi:10.1088/0264-9381/31/8/085016
[arXiv:1402.0657 [gr-qc]].

\bibitem{Duval:2014uva}
C.~Duval, G.~W.~Gibbons and P.~A.~Horvathy,
``Conformal Carroll groups and BMS symmetry,''
Class. Quant. Grav. \textbf{31} (2014), 092001
doi:10.1088/0264-9381/31/9/092001
[arXiv:1402.5894 [gr-qc]].

\bibitem{Duval:2014lpa}
C.~Duval, G.~W.~Gibbons and P.~A.~Horvathy,
``Conformal Carroll groups,''
J. Phys. A \textbf{47} (2014) no.33, 335204
doi:10.1088/1751-8113/47/33/335204
[arXiv:1403.4213 [hep-th]].

\bibitem{Hartong:2015xda}
J.~Hartong,
``Gauging the Carroll Algebra and Ultra-Relativistic Gravity,''
JHEP \textbf{08} (2015), 069
doi:10.1007/JHEP08(2015)069
[arXiv:1505.05011 [hep-th]].

		\bibitem{Bergshoeff:2017btm}
E.~Bergshoeff, J.~Gomis, B.~Rollier, J.~Rosseel and T.~ter Veldhuis,
``Carroll versus Galilei Gravity,''
JHEP \textbf{03} (2017), 165
doi:10.1007/JHEP03(2017)165
[arXiv:1701.06156 [hep-th]].
			
\bibitem{Guerrieri:2021cdz}
A.~Guerrieri and R.~F.~Sobreiro,
``Carroll limit of four-dimensional gravity theories in the first order formalism,''
Class. Quant. Grav. \textbf{38} (2021) no.24, 245003
doi:10.1088/1361-6382/ac345f
[arXiv:2107.10129 [gr-qc]].
			
			\bibitem{Andringa:2010it}
			R.~Andringa, E.~Bergshoeff, S.~Panda and M.~de Roo,
			``Newtonian Gravity and the Bargmann Algebra,''
			Class. Quant. Grav. \textbf{28}, 105011 (2011)
			
			\bibitem{Hansen:2021fxi}
            D.~Hansen, N.~A.~Obers, G.~Oling and B.~T. ~Sogaard,
            ``Carroll Expansion of General Relativity,''
        [arXiv:2112.12684 [hep-th]].
			
			
\bibitem{Bergshoeff:2018vfn}
E.~A.~Bergshoeff, K.~T.~Grosvenor, C.~Simsek and Z.~Yan,
``An Action for Extended String Newton-Cartan Gravity,''
JHEP \textbf{01} (2019), 178
doi:10.1007/JHEP01(2019)178
[arXiv:1810.09387 [hep-th]].

\bibitem{Bergshoeff:2020fiz}
E.~Bergshoeff, J.~Gomis and P.~Salgado-Rebolledo,
``Non-relativistic limits and three-dimensional coadjoint Poincar\'e gravity,''
Proc. Roy. Soc. Lond. A \textbf{476}, no.2240, 20200106 (2020)

\bibitem{Gwak:2015vfb}
S.~Gwak, E.~Joung, K.~Mkrtchyan and S.~J.~Rey,
``Rainbow Valley of Colored (Anti) de Sitter Gravity in Three Dimensions,''
JHEP \textbf{04}, 055 (2016)

\bibitem{Gwak:2015jdo}
S.~Gwak, E.~Joung, K.~Mkrtchyan and S.~J.~Rey,
``Rainbow vacua of colored higher-spin (A)dS$_{3}$ gravity,''
JHEP \textbf{05}, 150 (2016)

			
		\end{thebibliography}
	\end{document}